# Twistronics: A turning point in 2D quantum materials

Zachariah Hennighausen[1] and Swastik Kar[2]

[1] NRC Postdoc Residing at the Materials Science and Technology Division, United States Naval Research Laboratory, Washington DC 20375, USA
[2] Department of Physics, Northeastern University, Boston MA, 02115

## Abstract

Moiré superlattices - *periodic orbital overlaps and lattice-reconstruction between sites of high atomic registry in vertically-stacked 2D layered materials* - are quantum-active interfaces where non-trivial quantum phases on novel phenomena can emerge from geometric arrangements of 2D materials, which are not intrinsic to the parent materials. Unexpected distortions in band-structure and topology lead to long-range correlations, charge-ordering, and several other fascinating quantum phenomena hidden within the physical space *between* the (similar or dissimilar) parent materials. Stacking, twisting, gate-modulating, and optically-exciting these superlattices open up a new field for seamlessly exploring physics from the *weak* to *strong* correlations limit within a many-body and topological framework. It is impossible to capture it all, and the aim of this review is to highlight some of the important recent developments in synthesis, experiments, and potential applications of these materials.

## 1. Introduction

The field of Two-Dimensional (2D) Materials has evolved significantly since graphene's unique properties were first demonstrated in 2004,[1] and the Nobel Prize awarded in 2010. Graphene is largely considered the first 2D material to be isolated. Hundreds of stable monolayer materials have been predicted,[2–4] and tens of monolayer materials have been synthesized that span a range of different chemistries and crystal structures. This includes hexagonal boron nitride,[5–7] transition metal dichalcogenides (TMDs),[7,8] *X*enes (e.g., *X* = Si, Ge, Sn,…),[9,10] *MX*enes,[9,11–13] and Chromium Trihalides.[14–16] Additionally, an emerging field of quasi-2D materials, such as $Fe_3GeTe_2$, $Ni_3GeTe_2$, and perovskites, have shown promise to be stable in monolayer morphology.[17–19] Monolayer materials have made contributions to fields as diverse as sensors (optical, bio, chemical, strain),[20–26] spin current injection,[27] massless resonators,[28] neural interfaces,[29,30] neural scaffolds,[31,32] single-ion detection,[22] DNA/RNA sequencing and protein characterization,[25] low-cost organic solar cells,[33] single-photon emitters,[34] LEDs,[35–37] superconductivity,[38,39] transistors,[40–42] piezoelectrics,[43] and magnetism.[44]

Although a definitive definition of 2D materials has not been established, most 2D materials are few-atom-thick and their properties can be described by quasiparticle excitations confined in two dimensions (*i.e.* their band structure has two momentum dimensions, $k_x$ and $k_y$, and one energy dimension, with the $n^{th}$ band described by $E_n = E_n(k_x, k_y)$), and chemically, their surface bonds are completely satisfied (*i.e.* they do not have unsatisfied surface dangling bonds). This latter fact leads to the generalization that monolayers of these materials are characterized with strong in-plane bonding, and weak out-of-plane bonding results. This led to the concept of a new type of pseudo-crystalline solid obtained by vertical stacking of 2D materials, popularly known as van der Waals vdW layered solids.[45]

After studying monolayer materials in their pristine form, the next natural step was to vertically stack them to create 2D homo- and heterostructures. However, it was soon evident that such a 2D heterostructure's properties are not just a physical superposition of those of their individual layers, but are a complex outcome of the interlayer coupling, lattice reconstruction and twist angle, which can manifest and spawn unexpected new phenomena. (*Note*: *the stacking order, i.e. the manner in which the lattice points are stacked on top of each other even in untwisted self-similar 2D layers, could also play a role in the properties of the resulting heterostructures, but is intentionally not a topic covered in this review*). Indeed, the diverse variety of atomically-thin 2D materials provide unique opportunities for creating new nanostructures through vertical and horizontal stacking with virtually limitless possibilities and, as a result, the intense interest in exploring the science and engineering of such 2D structures has continued unabated since the discovery of graphene, with structurally-inspired names such as van der Waals solids or heterostructures, functionally-inspired names such as electron quantum metamaterials, [46–49] and with a unique area of focus being moiré superlattices,[50–52] which has revealed a diverse range of novel quantum phenomena and phases, often related to the specific way the layers are twisted with respect to each other. The twist angle (*i.e.*, the relative angle between layers)[53] determines the new periodicity (*i.e.*, moiré pattern) formed between the individual crystalline layers, while the interlayer coupling determines the magnitude of electron-particle interactions,[54] hybridization, charge redistribution,[55] and lattice reconstruction. Together, these factors induce the formation of a 2D moiré superlattice with a unique band structure that induces new properties, as follows.

When monolayer materials are vertically stacked, new properties only emerge if a sufficiently strong interlayer coupling is formed, which facilitates the hopping of electrons and redistribution of charge between the layers. More specifically, when the interlayer coupling is sharply reduced or severed, each layer becomes more independent, behaving similar to its monolayer configuration. Conversely, when the interlayer coupling is made strong, each layer's independence is reduced, and a superlattice forms that induces new properties. Stated simply, the interlayer coupling can facilitate the emergence of a new long-range periodicity, unique band structure, and dramatically new properties.

Molybdenum disulfide ($MoS_2$) is a layered vdW material, and demonstrates this concept well. There are notable changes to the band structure and properties between mono- and bilayer configurations. The monolayer's bright photoluminescence is significantly reduced in the bilayer configuration, due to the bandgap shifting from being direct to indirect.[56,57] As the layer count is increased, the band structure continues to evolve and approach that of the bulk, which is not photoluminescent.[56,57] However, the photoluminescence can be recovered in bulk $MoS_2$ by intercalating it with compounds that disrupt the interlayer coupling, inducing each layer to behave more independent and monolayer.[58–60]

Controllably manipulating the interlayer coupling has spurred a variety of technologies, including high density information storage,[61–63] spin-current filtering,[16] superdense and ultrafast lithium storage,[64,65] tunable LEDs,[63] topological transport,[66] ultralow power computing,[67,68] oxygen sensors,[61] and superconductivity.[69–71] Numerous techniques have been demonstrated to manipulate the interlayer coupling, including nano-squeegee,[72] controlled intercalation,[58,61] annealing,[73] and pressure tuning.[74–77] These techniques are described in greater detail in Section 4.

Twistronics is the field of how the relative angle (or twist) between the layers of a vertically stacked 2D structure affect its properties.[53] This effect is unique to 2D materials, and has demonstrated the ability to dramatically alter the material's band structure and properties. In certain 2D structures, the effect is so transformative that the corresponding twist angles are described as "magic".[68–70,78] As of publication, the twist angle has been shown to manifest and manipulate a wide-range of properties, including superconductivity,[68–70,79] orbital ferromagnetism,[67,68] moiré excitons,[80–85] the quantum anomalous Hall effect,[86,87] strong electron-electron correlations,[69,88,89] both intra-[56,61,63] and interlayer[90,91] exciton photoluminescence, twist-dependent color,[92] Raman modes,[93–96] Hofstadter Butterfly,[97] 1D topological channels,[98] fractal quantum hall effect,[97,99] both lateral[98] and vertical conductivities,[100] and spatially-dependent vertical conductivity.[90,101]

The large body of literature suggests that the twist angle is as important as other major variables - such as doping, crystal structure, stoichiometry – and holds significant promise for next-gen technologies. In fact, differences in twist angle as small as 0.1° can induce transformative changes.[67,69,70,98] This naturally leads to the questions: How does the twist angle induce large changes to the band structure and properties in 2D materials, and why is it so sensitive? The answer, briefly stated, is due to the moiré pattern produced by the interaction of the layers, which induces the formation of a new periodic superlattice that has a unique band structure. With this in mind, the moiré pattern (and moiré superlattice) is strongly influenced by atomic reconstruction, which is when the atoms shift position due to an interplay between inter- and intralayer bonding. More specifically, depending on the specific 2D structure and the twist angle, the interlayer bonding induces a force that leads a repositioning of the atoms, where the equilibrium configuration is a balance between the inter- and intralayer bonding. This effect is only present when the interlayer coupling is sufficiently strong to enable sufficiently robust interlayer bonding. Further, the moiré pattern evolves continuously with twist angle, suggesting the superlattice – and corresponding band structure - evolve as well.

This review article is organized into five sections, as follows:

1. Introduction

2. Theory of Twist-Dependent Effects

3. Twist-Dependent Experimental Discoveries

4. Twisted 2D Structure Fabrication Methods

5. Future Outlook

## 2. Theory of Twist-Dependent Effects

### 2.1. Twist-angle determines the periodic moiré superlattice

When two periodic functions are effectively multiplied – even two simple cosine functions - new periodicities emerge.[102–106] Similarly, when two lattices (or crystals) are effectively multiplied, a new periodic lattice emerges called a "moiré pattern". The concept of a moiré pattern – and the formation of a moiré superlattice – is central to understanding the origin of twist-dependent properties.

The etymology of the word moiré is French, and refers to two layers of watered silk pressed together. As the layers slide across each other, an interference pattern is created. Similarly, as two lattices slide across each other, moiré patterns are created.

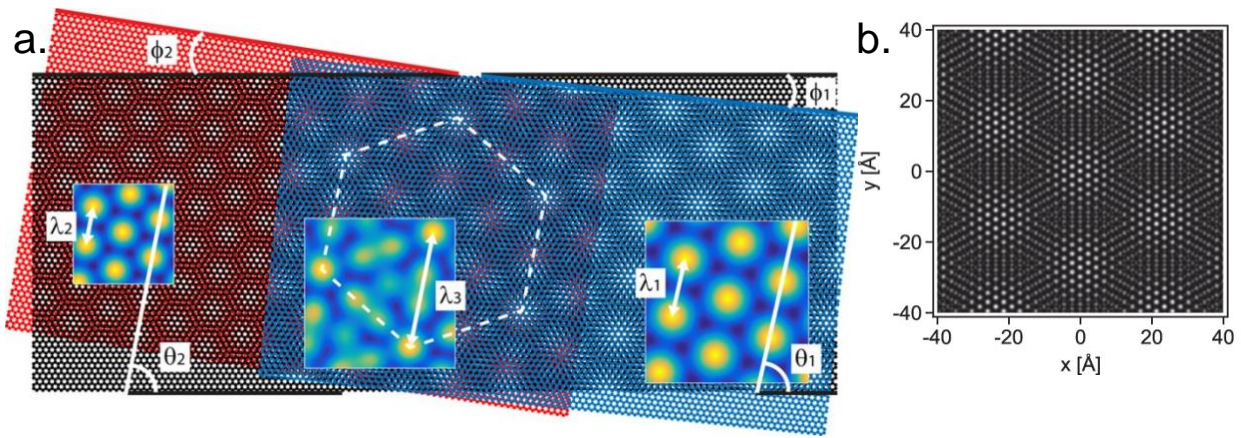

**Figure 1: Moiré patterns generated from overlaying hexagonal lattices.** (a) Three different lattices (red, black, and blue) consecutively superimposed in that order, where they represent hBN, graphene, and hBN, respectively. hBN and graphene have the same crystal structure, but the hBN in-plane lattice constant is 1.8% larger. The λ values show the size of the moiré pattern. Note, the moiré pattern changes with twist angle. (b) Plot of $f_a(\vec{r}) \times f_b(\vec{r})$ (see Eq. 1) with a twist angle of 0° and a relative lattice constant of ~1.1, demonstrating that moiré patterns can be fully described using a mathematical formulism. Reprinted (a) and (b) with permission from [107] and [105], respectively.

Figure 1 demonstrates four different moiré patterns using hexagonal lattices twisted to a variety of angles, and then overlaid (superimposed). When images are superimposed, they are effectively multiplied because the pixels are overwritten, vice summed, thereby introducing non-linear terms. Figure 1a shows three different hexagonal lattices, representing hexagonal boron nitride (hBN), graphene, and hBN, respectively, thereby forming an hBN/graphene/hBN 2D heterostructure.[107] As the twist angle is rotated, the moiré pattern evolves, changing its relative twist angle and size. The λ value quantifies the moiré pattern size. Using Eq. 1, Figure 1b is a plot of $f_a(\vec{r}) \times f_b(\vec{r})$ with a twist of 0° and a lattice constant ratio equal to ~1.1, demonstrating a mathematical formulism to describe moiré patterns.[105] The twist angle can be specified using an appropriately chosen set of $\vec{k}$ vectors.

$$f(\vec{r}) = \frac{1}{9} + \frac{8}{9}\cos(\frac{1}{2}\vec{k_1}\vec{r})\cos(\frac{1}{2}\vec{k_2}\vec{r})\cos(\frac{1}{2}\vec{k_3}\vec{r})$$

$$\text{with } \vec{k_1} = k(1,0);\ \vec{k_2} = k(\bar{1},1);\ \vec{k_3} = k(0,\bar{1});\ and\ k = \frac{2\pi}{a\sqrt{3}/2} \tag{1}$$

The moiré pattern forms a new long-range periodicity that can be described as a commensurate unit cell with its own vectors. These vectors are then used to identify and specify the new superlattice (or supercell) formed. Eq. 2 shows the Diophantine equation that must be satisfied to form a commensurate unit cell, where $m$, $n$, $r$, and $s$ must be integers, and $\vec{b}$ and $\vec{c}$ are the lattice vectors for each parent material. Since the lattice vectors include information about the lattice constant and orientation in space, the moiré superlattice formed is dependent on both the ratio of lattice constants and the twist angle. The following references contain amplifying information on mathematically constructing moiré patterns for both theory and experiment 2D materials applications, including identifying grain boundaries, and probing possible room temperature Wigner crystals existing in the interlayer region.[55,103–106,108,109]

$$\vec{v}_{moiré} = m \times \vec{b}_1 + n \times \vec{b}_2 = r \times \vec{c}_1 + s \times \vec{c}_1 \tag{2}$$

### 2.2. Modeling the interlayer coupling and moiré superlattice

As discussed in the introduction, the band structure and properties of a 2D material can shift dramatically between mono- and bilayer configurations, if the interlayer coupling is sufficiently strong. The interlayer coupling facilitates the movement of electrons and redistribution charge between the layers, reducing each layer's independence, and forming a distinct material. The interlayer coupling facilitates the hopping of electrons and redistribution of charge between the layers, enabling electrons to "perceive" both layers (or lattices) simultaneously, inducing the formation of a new periodic moiré superlattice and corresponding band structure. This knowledge is critical for understanding twist-dependent effects, and is summarized as follows:

*Twist-dependent effects are a product of the interaction between the layers. As such, a sufficiently strong interlayer coupling is required to detect twist-dependent effects.*

Understandably, the earliest theoretical treatise of vertically stacked twisted layers were performed on graphene. In 2007, one of the earliest electronic structure calculations applied to a twisted stacked bilayer graphene system used model Hamiltonians for a continuum electronic description in terms of massless Dirac fermions, coupled by a slowly varying periodic interlayer hopping,[110] that showed the slowing-down of the Fermi velocity. Further, tight-binding (in 2010)[111] and continuum model Hamiltonian calculations (in 2011) revealed nondispersive flat bands close to the Fermi energy for very specific small-angle twists.[78] Following the recent observation of unconventional superconductivity below 1.7K in twisted bilayer graphene at a twist angle θ ~1.1°, extensive work has been initiated to theoretically model the interlayer coupling and moiré superlattice in 2D materials, where methods are often tailored to certain applications.[112–117]

In beyond-'magic-angle' twistronics, especially for larger angles with smaller moiré-superlattices, First-principles density functional theory techniques provide a powerful advancement. Recent Density Functional Theory (DFT) work using the strongly constrained appropriately normed (SCAN) functional[118] have demonstrated impressive success in modeling monolayer materials.[119] This is an especially noteworthy accomplishment because SCAN is the only known meta-GGA approximation that is fully constrained, obeying all 17 known exact constrains. Said another way, SCAN is arguably the most realistic and accurate modeling approach developed for condensed matter systems to date. It is likely that SCAN will play a growing role in synergy between theory and experiment, including better understanding twistronics experimental results, and predicting new twistronics materials and capabilities to guide experimentalists. Despite its success, it is relatively costly to implement, scaling approximately cubically to the number of atoms in the superlattice (i.e., $n^3$).

With this in mind, the theoretical community developed advanced methods that rely on assumptions to reduce the computational cost. One of the first publications targeting twistronics applications showed fast and efficient simulation of multilayered stacks in the presence of local disorder and external fields.[53] Despite the model's simplicity and lower cost, it matches certain complex experimental results. More specifically, it correctly calculates the quantization of Hall conductivity in twisted bilayer graphene in the presence of magnetic fields, and the twist-dependent band gap in certain TMDs. While SCAN may be limited to several hundred atoms for practical calculations, this method accommodates over one million, enabling massive moiré superlattices to be studied.

Previous work has demonstrated atomic reconstruction in vertically-stacked 2D structures, where the interlayer coupling forces and moves atoms into a new configuration. Modeling atomistic simulations of superlattices in 2D materials is possible, but very expensive due to the long range of interlayer interactions. A theory method demonstrated success and agreement with experiment using a nonlinear finite element plate model with the interactions between layers described by an efficient and accurate discrete-continuum interlayer potential.[120] This method used Kolmogorov-Crespi model to simulate the interlayer coupling, a popular technique to capture interlayer graphene interactions dating back to 2005.[121] The technique showed that certain exchange-correlation functionals are able to partially capture the interlayer coupling. More specifically, the generalized gradient approximation (GGA) and the local density approximation (LDA) partially capture the interlayer interaction, despite the fact that GGA and LDA are in qualitative disagreement. Additionally, the Lennard-Jones potential does not capture the effect, suggesting robust exchange-correlations are required, and the interlayer coupling is not a simple chemical potential. An updated Kolmogorov-Crespi model for TMDs has been proposed.[122]

It has been shown that the real-space misalignment of two repetitive patterns leads to localized states in a small region of momentum space, allowing quantum mechanics to be re-formulated, where the underlying two-dimensional lattices no longer explicitly appear, enabling more efficient calculations.[109]

## 3. Twist-dependent experimental discoveries

This chapter aims to cover the most important experimental discoveries in the field of twistronics. As of publication, a vast majority of the discoveries in twistronics are using twisted bilayer graphene (tBLG) or other graphene-based 2D structures (e.g., graphene+hBN). This is likely due to two factors: First, the unique interaction graphene can have with other materials due to its out-of-plane Π bonds, which are predicted to significantly affect the interlayer coupling.[123][124] Second, the interlayer spacing (or distance) makes up a considerable volume in tBLG 2D structures, suggesting the interlayer interaction is proportionally more influential.

There are a handful of experimental publications about non-graphene 2D structures, suggesting the field has significant room for growth. Atomic reconstruction was demonstrated in twisted TMD/TMD 2D structures.[90] Evidence of a purely charge lattice that resides in the interlayer region was demonstrated using $Bi_2Se_3$/TMD 2D structures.[55]

### 3.1. Influence of the interlayer region

The interlayer layer region (volume) plays a central role in vertically-stacked vdW and twisted 2D structures. Discussed in greater detail next, the interlayer region separation in 2D structures is likely between 0.3-0.4nm, a size that is greater than the diameter of most atoms, but close to the in-plane atomic bonding of prominent 2D materials (e.g., graphene, TMDs). This comparatively large size suggests that a large fractional volume of the 2D structure is the interlayer region. In summary:

*The 2D structure's properties are significantly influenced by both the interlayer region's size and what comprises it (e.g., vacuum, charge, electrons, atoms).*

As mentioned previously, and discussed in more detail later, the interlayer coupling plays a critical role in manifesting new properties. This naturally leads to the question: How could a strong interlayer coupling form across a large interlayer separation that is van der Waals bonding in bulk morphology?

Initially this may not be a surprising question because the interlayer distance (i.e., 0.3-0.4nm) is close to the in-plane bonding lengths of prominent 2D materials (e.g., graphene is 2.46 Å). If power-law hopping strengths are assumed, the hopping parameters may not deviate significantly. However, hopping parameters change depending on the bonding. For example, sigma bonding in graphene contributes to the high tensile strength and likely facilitates a strong in-plane conduction compared to out-of-plane.

Previous work showed that electron tunneling across the interlayer region is likely present and affects the properties,[125] some publications demonstrate minimal contribution therefrom,[126] suggesting other effects are influencing the interlayer coupling. Previous publications use DFT calculations and experiment to show that the interlayer coupling likely induces significant charge redistribution into the interlayer region, thereby modifying the hopping parameter, and facilitating electron transfer and communication between the layers.[55,62]

### 3.1.1. Interlayer region comprises a large fractional volume

In tBLG 2D structures, approximately one third of the volume is taken by the interlayer region, thereby bestowing it tremendous influence over the charge distribution and band structure. Graphene layers are predicted to be ~0.4nm tall,[127,128] while the interlayer spacing is very likely between 0.3-0.4nm,[129] in close agreement with graphite (0.34nm).[130] These values are in agreement with numerous robust theory calculations.[121,131–134]

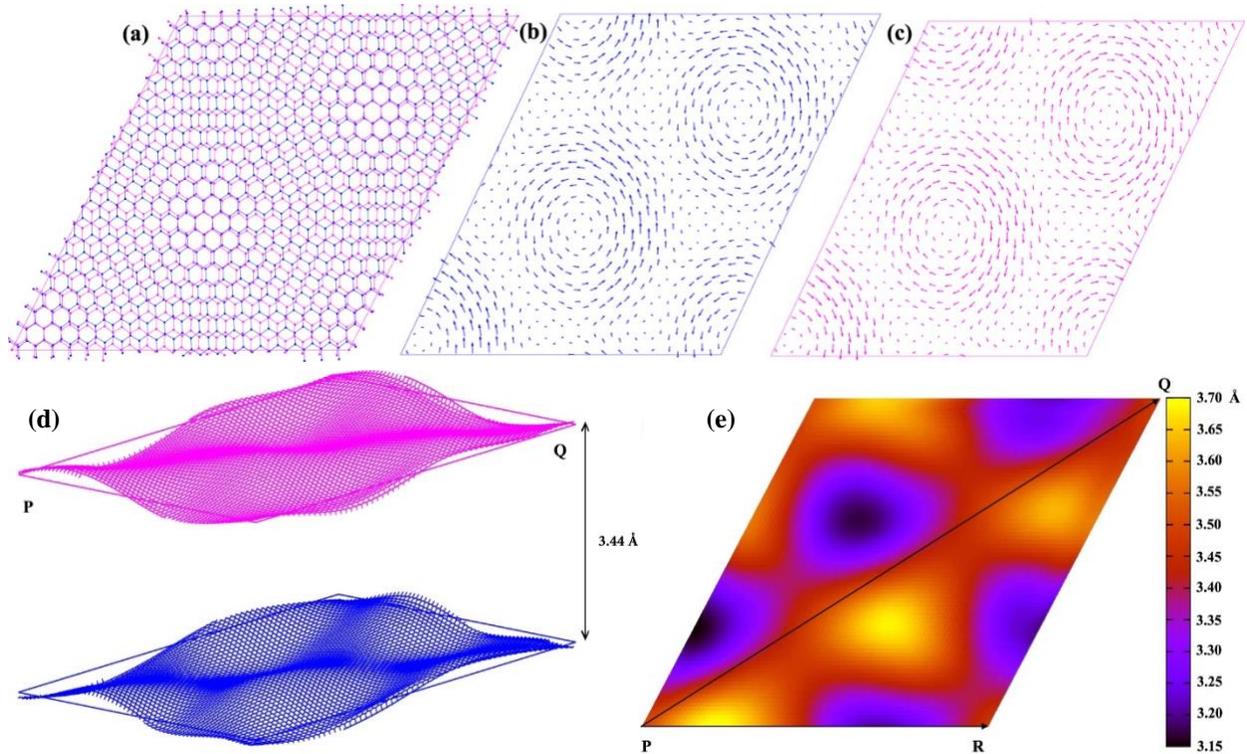

**Figure 2: Effect of interlayer coupling on twisted bilayer graphene.** (a)-(c) Simulation of twisted bilayer graphene (tBLG) twisted to 5.09°. (a) Before reconstruction. (b) Bottom layer after reconstruction where atoms rotate counterclockwise near the vortex. (c) Top layer after reconstruction where atoms rotate clockwise near the vortex. (d)-(e) Buckling in tBLG twisted to 1.61° using 15132 atoms. The equilibrium separation between both layers is calculated to be 3.44 Å. (e) Buckling profile of the top layer. The buckling height is 0.51 Å. Reprinted all images and portions of the caption with permission from [129].

Previous experimental work suggests that the interlayer spacing is dependent not only on the twist angle,[135] but also variations within the moiré superlattice as certain areas are raised or buckled.[120,124] The raising and buckling, shown in Figure 2, is due to atomic reconstruction from interlayer forces shifting the energy minima of the atoms (discussed further in Section 3.3). To the best of our knowledge, definitive measurements of the interlayer distance in bilayer graphene are still ongoing and of high interest to the community. This is in part due to the important role the interlayer separation has in coupling the layers. Previous experimental work demonstrated that the interlayer coupling is strengthened when the layers are hydrostatically forced together,[74,75,79] or when an AFM presses the layers together after cleaning the interlayer reggion.[72] These experiments are discussed in greater detail later. The smaller the interlayer spacing, the more likely electrons can hop between layers.[136]

### 3.2. Significant charge redistribution into the interlayer region

A reoccurring and central topic of this review article is the interlayer coupling in 2D structures, a foundational concept arising from quantum mechanical interactions that manifest unique properties not observed in either layer individually. Throughout the article we discuss a large body of experimental work demonstrating the significant effect the interlayer coupling has on the properties, methods to controllably tune it, and promising technological applications. Hence, there are clear differences between 2D structures with strong *vs.* weak interlayer coupling, suggesting the mechanisms at a quantum mechanical level are made stronger or weaker, respectively. This naturally leads to the following central questions:

> *What are the physical changes that describe the interlayer coupling (e.g., charge redistribution, atomic reconstruction)? And how do these physical changes facilitate new properties?*

Additionally, how does the interlayer coupling exert such a large influence across a relatively large interlayer separation (~0.3nm)? This is made even more surprising when one considers that the interlayer bonding in their bulk counterparts is considered to be relatively weak and van der Waals.

Despite its fundamental importance, significant questions remain regarding the physical origin and mechanisms that facilitate the interlayer coupling. This is in part due to the difficulty of directly probing the interlayer region using an experiment. For example, scanning tunneling microscope (STM) measurements have produced impressive strides in this area, but do not probe the interlayer region directly. They probe the local density of states (LDS) with sub-angstrom spatial resolution, but are limited in that the measured values are a combination of both the interlayer coupling, as well as the individual top- and bottom-layer atoms. Although the interlayer coupling's influence on different properties can be probed (e.g., twist-dependent friction,[137,138] excitons,[58,61] Raman modes[93,96]), none of these experimental setups directly probe the interlayer region.

Two techniques have demonstrated success in directly probing the interlayer region: X-ray diffraction, [139] and high-energy electron beam diffraction.[55] X-Ray diffraction is able to probe the electron density (ED) of a crystal, when using synchrotron sources. The exceptionally small wavelengths (~0.0025nm) of the high-energy (200keV) electrons enable each layer, as well as the interlayer region, to be independently probed. More specifically, the interlayer separation is approximately ×40 the electrons' wavelength, suggesting interference is impossible between the layers. Together, this enables scattering sites in between the layers to be independently probed. It should be noted that robust density functional theory (DFT) calculations are very likely required to fully interpret the experimental results and make strong claims about the interlayer region.

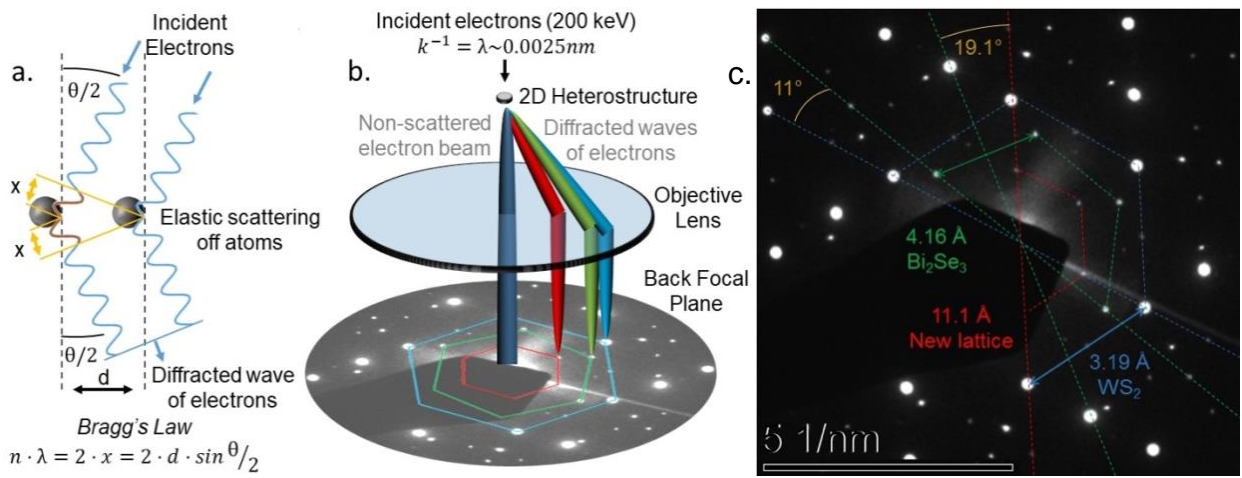

**Figure 3: Simplified descriptions of pertinent SAED concepts.** (a) Scattering mechanism for high-energy selected area electron diffraction (SAED) follows Bragg's law, where electrons incident at a non-zero angle elastically scatter from atoms. In the diagram shown, the left electron must travel a longer distance (i.e., 2·x) for a diffracted wave front to form. In sharp contrast to multi-slit diffraction optical setups, the electrons must both elastically scatter and be incident at a non-zero angle for diffraction to occur. (b) Simplified diagram of a TEM in SAED mode. Note, each lattice diffracts at a different angle because their lattice spacing's are different. (c) SAED image of a $Bi_2Se_3$/$WS_2$ 2D heterostructure with the in-plane lattice parameters labeled. The distinct dots (labelled in blue and green) indicate the TMD and $Bi_2Se_3$, respectively, both grow highly crystalline. A third, new set of diffraction spots (red label) is present that has no known atomic basis and follows the geometric moiré superlattice pattern of the 2D heterostructure. Moiré patterns only manifest when lattices are effectively multiplied,[102–106] suggesting the strong interlayer coupling induces significant charge redistribution into periodic electron scattering sites. It is unlikely that wave-interference or double diffraction are able to produce moiré superlattice spots in high-energy SAED. Reprinted images and portions of the caption with permission from [55].

Figure 3 shows a simplified diagram of selected area electron diffraction (SAED), along with experimental results demonstrating the presence of a lattice (i.e., a periodic array of scattering sites) with no known atomic basis. Figure 3a shows Bragg's law, which is when particles with wave-particle duality elastically scatter from an array of sites, enabling wave fronts with constructive interference to form bright spots at the detector. Note, the angle of the diffracted wave front (i.e., θ) is dependent on the spacing between the scattering sites. Figure 3b shows how each wave front travels independent of the others, suggesting interference of the wave fronts is unlikely. In summary:

*The presence of bright spots in SAED images indicates with a high degree of confidence that constructive wave fronts are forming due to a periodic array of scattering sites.*

Figure 3c shows an experimentally obtained SAED image of a $Bi_2Se_3$/$WS_2$ 2D heterostructure with the corresponding lattice constants labeled. Note, a large (1.11nm) lattice is clearly present that does not correspond to either parent material. This lattice matches the moiré superlattice nearly perfectly (see Figure 4c), suggesting that the new lattice is a product of the interlayer coupling. As discussed in the theory section, moiré patterns and moiré superlattices only form when there is an effective multiplication between the two materials.[102–106]

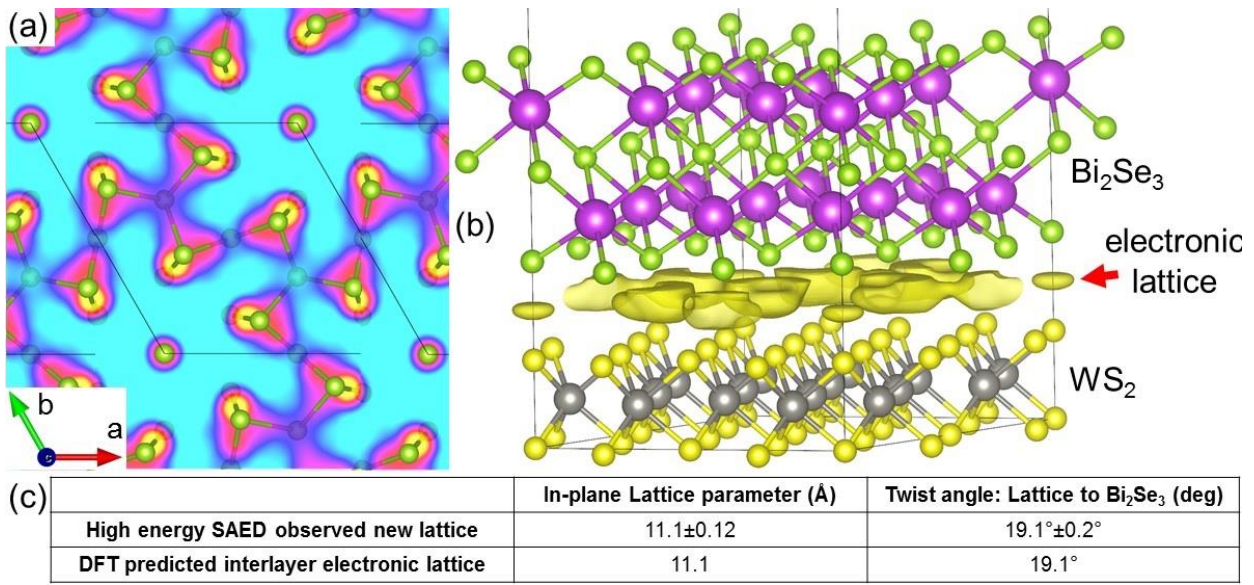

**Figure 4: Evidence of a purely electronic two-dimensional lattice at the interface.** (a) DFT calculated interlayer charge redistribution due to the interlayer coupling of a twisted $Bi_2Se_3$/$WS_2$ 2D heterostructure, using experimentally measured lattice parameters. The image is a cut of the plane lying equidistant from both materials, with the nearest neighbor atoms included. Note, the charge redistribution is concentrated between nearest interlayer neighbors. (b) Cross-section view for the same DFT calculated 2D heterostructure, showing charge pools form between nearest neighbors. Since the atomic registry (or nearest neighbors) is determined by the twist and moiré pattern, the charge pools form a purely electronic lattice that follows the moiré pattern. (c) Table with the parameters from the experimentally measured third-crystal and the DFT calculated interlayer electronic lattice, demonstrating their complete agreement. Together, the experimental and theoretical work suggest that the high energy electrons might be scattering from the DFT-predicted interlayer charge pools. Reprinted images and portions of the caption with permission from [55].

It is highly unlikely that the moiré superlattice diffraction spots result from either interference of the parent layer diffraction patterns, or the presence of foreign atoms, suggesting that a new non-atomic lattice composed entirely of charge (*i.e.*, purely electronic) is present. This conclusion naturally leads to the following questions: (1) What do the electronic moiré superlattice scattering sites look like, (2) where do they reside, and (3) why does the lattice follow the moiré pattern, vice a different structure?

DFT calculations, which previously demonstrated success predicting structural and electronic information in solids,[140] were accomplished to elucidate answers (see Figure 4). The synergy between DFT-theory and experiment have previously demonstrated success probing the charge distribution in solids.[39,62,139,141–143] Figure 4 shows first principles DFT calculations that predict the formation of a purely electronic 2D lattice residing in the interlayer region whose unit cell size and relative orientation are in agreement with the experimentally observed moiré pattern from Figure 3, suggesting that such a purely electronic lattice may be responsible for the experimentally observed diffraction spots. The 2D electronic lattice has concentrated pools of charge located at sites of nearest-neighbor atomic registry. Previous work found evidence that strong electronic correlations play a role in the atomic reconstruction and charge redistribution using a synergy between robust transport experiment and theory.[89]

### 3.3. Atomic reconstruction

The twist angle not only affects the interlayer spacing, but can also induce atomic reconstruction at the vdW interface, a process where the atoms strain themselves in-plane to accommodate out-of-plane interlayer forces.[98,129,144] Said another way, the interlayer forces alter the energy landscape and move atoms into energetically more favorable positions. This process can induce significant buckling (see Figure 2), and has been shown to introduce drastic changes, such as scattering high-energy electrons and shifting the electronic properties.[98] The experimental results are in agreement with theory predictions.[55,120,144–146]

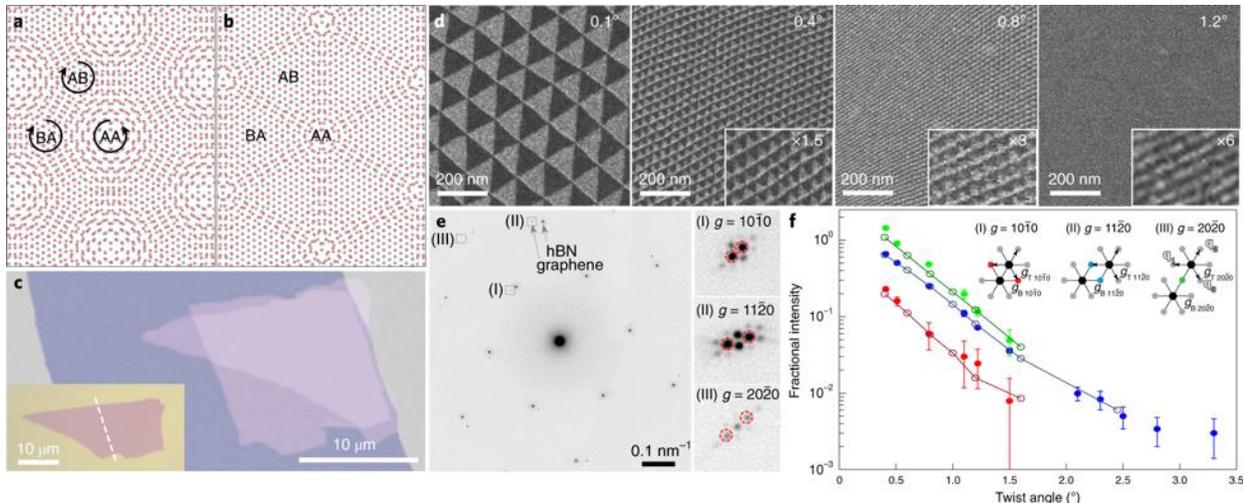

**Figure 5: Atomic reconstruction in twisted bilayer graphene.** (a)-(b) Schematic diagrams of tBLG (a) before and (b) after atomic reconstruction. (c) Optical microscope image of tBLG covered with an hBN layer, where the inset shows the original graphene layer, and the dashed line indicates the fold boundary. (d) TEM dark-field images of the graphene diffraction peak (g=10$\bar{1}$0) in tBLG twisted at various angles. The alternating AB/BA domains are visible. (e) SAED of tBLG (twisted to 0.4°) and hBN 2D heterostructure, with corresponding peaks labeled. (f) Plot of fractional intensity ($I_{sat}/I_{Bra}$) as a function of twist angle, where the filled circles are from experiment, while the open circles and line fits were obtained from simulated SAED patterns. Amplifying information on the simulated SAED patterns are found in the following reference [120]. The red, blue and green colors correspond to the satellite peaks. Bragg peaks of graphene are represented as black filled circles, and satellite peaks are represented as red, blue, green and grey circles. Reprinted images and portions of the caption with permission from [98].

Figure 5 shows, to the best of the authors' knowledge, the first experimental confirmation of reconstruction in tBLG. Yoo *et al* used the synergy of numerous techniques (i.e., selected area electron diffraction (SAED), dark-field TEM (DF-TEM), multislice simulations, and transport measurements) to show there is both atomic and electronic reconstruction at the vdW interface in tBLG.[98] Multiple tBLG 2D structures were prepared at various twist angles near 0°. New SAED diffraction spots appeared, whose intensity varied depending on the twist angle, an indication of the magnitude of the interlayer coupling. Using DF-TEM, changes in the electron-scattering properties were demonstrated over domains tens to hundreds of nanometers in size. It should be noted that double diffraction is relatively uncommon and unlikely in TEM measurements,[147,148] suggesting that both the new SAED spots, and the DF-TEM contrast changes, are primarily a product of the interlayer coupling and interaction. These discoveries reinforce the incredible finding made in 2018 that the electrons in tBLG can exhibit strongly correlated behavior and band flattening due a sufficiently intense interlayer coupling.[69,70] Theoretical work predicting correlated electron behavior was completed approximately seven years earlier in 2011.[78] The SAED measurements are in agreement with robust multiscale simulations that model such diffraction.[120]

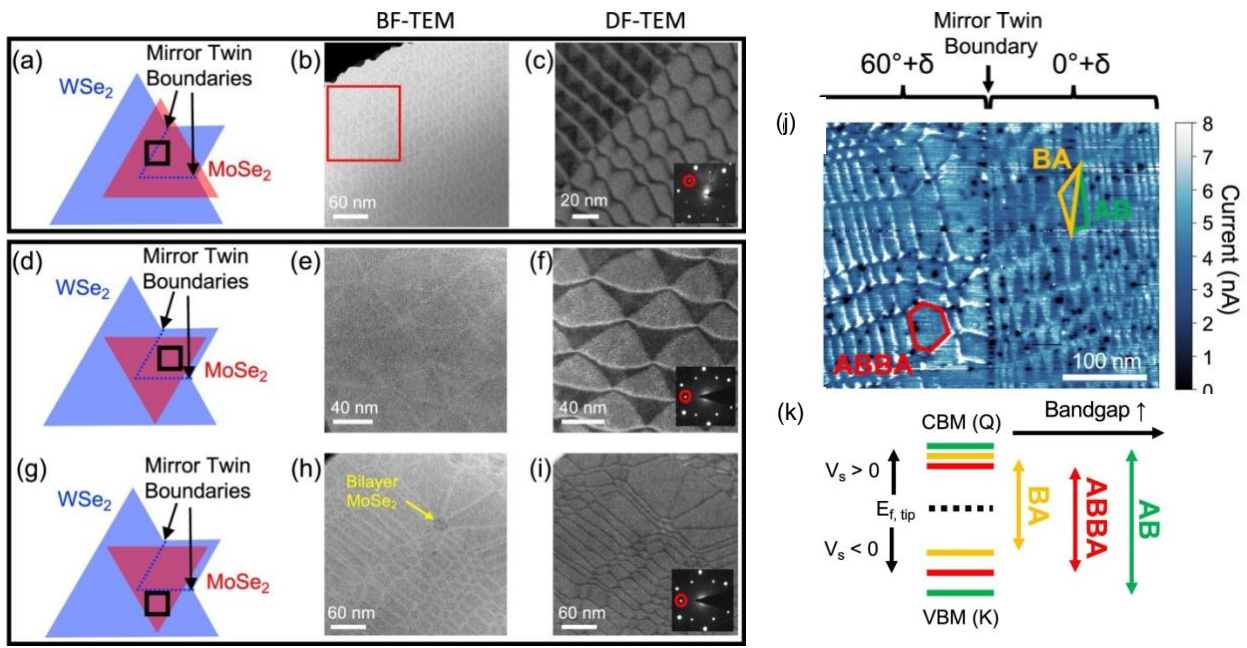

**Figure 6: Atomic reconstruction in twisted bilayer TMD 2D heterostructures.** (a)-(i) TEM measurements of 1L $MoSe_2$/1L $WSe_2$ 2D heterostructures at different twist angles. (a),(d),(g) Schematics for each sample where a monolayer $MoSe_2$ triangle was mechanically transferred on top of a $WSe_2$ monolayer with mirror twin boundaries, resulting in both 0°+δ and 60°+δ heterostructures in the same sample with the same misalignment angle, δ. The black boxes in all schematics indicate the approximate location of TEM images. (b),(e),(h) Bright Field-TEM images. (c),(f),(i) Dark Field-TEM images. Note that clear grain boundaries are detected, along with hexagon-like domains, suggesting the interlayer coupling is facilitating the formation of a moiré superlattice. (j) Conductive-AFM mapping measurement of a 1L $MoSe_2$/1L $WSe_2$ 2D heterostructure with a voltage bias applied to shift the Fermi energy and tip-sample voltage bias. Similar to the TEM measurements, the domains can be clearly discerned, demonstrating the sharp spatial variation in conductivity. (k) DFT calculated valence band maxima (VBM), conduction band minima (CBM), and the band gap for the AB, BA, and ABBA crystal structures. Both the CBM and the VBM of BA lie within the band edges of AB, consistent with the gold domain labeled in (j) being more conductive than the green domain. Reprinted images and portions of the caption with permission from [90].

The effect of atomic reconstruction has been demonstrated in bilayer TMD-based 2D heterostructures, including $MoS_2/WS_2$ and $MoSe_2/WSe_2$ 2D heterostructures, where both DF-TEM measurements and conductive-AFM mapping measurements show clear boundaries and hexagon-like features, suggesting the interlayer coupling is inducing the formation of a moiré superlattice.[90]

Work on $Bi_2Se_3$/TMD 2D heterostructures demonstrated that the moiré superlattice can produce a corresponding SAED diffraction pattern.[55] Using DFT calculations, the authors show that the likely origin and mechanism for the new spots is a pure charge crystal that exists in the interlayer region. Figure 5 shows tBLG SAED spots that likely have a similar origin as the moiré superlattice diffraction in $Bi_2Se_3$/TMD 2D heterostructures, suggesting scattering sites composed entirely of charge might be scattering high-energy electrons in tBLG. The presence of new SAED spots is a direct measurement of a periodic array of scattering sites. The SAED images show a new set of scattering sites that are distinct from the atoms in the parent layers.

Of note, moiré superlattice SAED spots have only been detected in two types of 2D structures (i.e., tBLG and $Bi_2Se_3$/TMD),[55,98] and have been noticeably absent in other 2D structures, including atomically reconstructed TMD 2D structures. Together, this suggests that the interlayer coupling might be strong enough in certain 2D structures to redistribute charge into new scattering sites.

The discoveries of atomic and electronic reconstruction in 2D materials demonstrated the outsize influence the interlayer coupling has on the crystal structure and charge distribution, which directly affects the band structure. This is important for the field of 2D materials going forward because simply analyzing the moiré pattern produced by two monolayer materials is not a realistic assumption about the system. Researchers must assume that the atoms, and corresponding moiré pattern, may have shifted and reconstructed when the materials were coupled together.

### 3.4. Twist-dependent transport: superconductivity at "magic" angles

2D Materials have made numerous contributions to the field of superconductivity. $MoS_2$ exhibits gate-induced superconductivity from bulk-like to individual monolayers,[149] as well as proximity-induced superconductivity in the monolayer configuration.[39] Amplifying information on superconductivity in 2D materials can be found in the following review article [150].

There is, however, one discovery that prominently stands out because it is twist-angle-dependent. Unconventional superconductivity was discovered in tBLG at certain "magic" angles likely due in part to band flattening.[69,70] The band-flattening-induced superconductivity is in part due to the strong electronic correlations and electron-electron interactions that manifest at small twist angles near 1°, and when the interlayer coupling is made sufficiently intense. The small twist angles produce a moiré superlattice that contains band flattening.[78] The strong correlations are sufficiently intense in tBLG to induce fractal quantum Hall states,[97] 1D topological insulating states,[66,98,151] and a Hofstadter butterfly.[89,97]

To best of the authors' knowledge, tBLG is the only material where twist-dependent superconductivity has been demonstrated, bestowing it a special status in both the superconductivity and the 2D materials communities. Other 2D materials have been predicted to have twist-dependent superconductivity,[152] but, as of publication, tBLG is the only known published experimental confirmation. Of note, previous work demonstrated that $Ca^+$ intercalated bilayer graphene is also superconducting till 4 K,[71,153] and theory predicts intercalating other alkali metals induces a similar effect.[154]

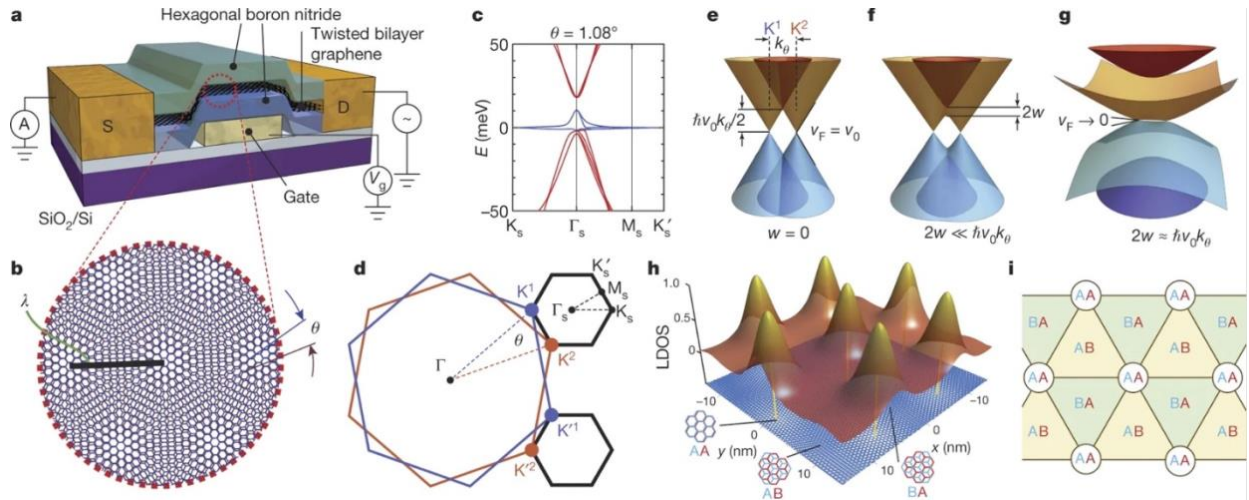

**Figure 7: Band flattening at "magic" angles in tBLG.** (a) Schematic of tBLG devices on SiO$_2$ substrates, where the tBLG is encapsulated in 10-30nm hexagonal boron nitride flakes. The conductance is measured with a voltage bias of 100 μV while varying the local bottom gate voltage V$_g$. 'S' and 'D' are the source and drain contacts, respectively. (b) The moiré pattern formed by tBLG. The wavelength is $\lambda = a/[2\sin(\theta/2)]$, where a = 0.246 nm is the lattice constant of graphene and θ is the twist angle. (c) The band energy $E$ of magic-angle (θ = 1.08°) tBLG calculated using an *ab initio* tight-binding method. The bands shown in blue are the flat bands that we study. (d) The mini Brillouin zone is constructed from the difference between the two K (or K`) wave vectors for the two layers. Hybridization occurs between Dirac cones within each valley, whereas intervalley processes are strongly suppressed. $K_S$, $M_S$, and $\Gamma_S$ denote points in the mini Brillouin zone. (e)-(g) Illustration of the effect of interlayer hybridization for (e) $w = 0$, (f) $2w \ll \hbar v_0 k_\theta$, and (g) $2w \approx \hbar v_0 k_\theta$; v$_0$ = 10$^6$ ms$^{-1}$ is the Fermi velocity of graphene. (h) Normalized local density of states (LDOS) calculated for the flat bands with $E > 0$ at θ = 1.08°. The electron density is strongly concentrated at the regions with AA stacking order, whereas it is mostly depleted at AB- and BA-stacked regions. (i) Top view of a simplified model of the stacking order. Reprinted images and portions of the caption with permission from [69].

The superconductivity observed in tBLG is described as "unconventional", in part because it cannot be explained by weak electron-phonon interactions.[70] The superconductivity only manifests at certain "magic" angles, which induce a significant change in the band structure that manifest a variety of notable effects. First, the bands from the electronic structure flatten, a process that enables electrons of the same energy to exist across a spectrum of momenta. Secondly, the bands exist near zero Fermi energy, resulting in a near zero Fermi velocity, and correlated insulating states at half-filling. Thirdly, a large effective mass for the electrons has been measured, indicative of the significant changes to the band structure.

Figure 7 discusses these concepts using theoretical calculations and diagrams reprinted from [69]. Figure 7a is a diagram of the devices constructed where the tBLG is encapsulated with hBN. Figure 7b is a representative moiré superlattice. Figure 7c shows the band structure for tBLG twisted to 1.08° using an *ab initio* tight-binding model previously demonstrated to be in agreement with experiment.[155] The calculations show a twist-dependence, in part due to the atomic field distortions of the out-of-plane Π (pz) orbitals.[155] Figure 7d shows how the low-energy band structure of tBLG can be considered as two sets of twisted Dirac cones, where the difference between the K and K` wavevectors manifests a moiré superlattice Brillouin zone. Importantly, interlayer hybridization enables the Dirac cones near the surface to interact, while more distant ones remain independent.[78,156] Figures 7(e)-(g) show the evolution of the band structure as the interlayer hybridization energy ($2w$) is increased, compared to the electron kinetic energy. The following statement is important for understanding how flat bands manifest due to the interlayer coupling:

> *The hybridized states evolve dependent on the ratio between the interlayer hybridization energy ($2w$) vs. the electron kinetic energy ($\hbar v_0 K_\theta$). More specifically, as the ratio is increased until it approaches 1 (i.e., when $2w \approx \hbar v_0 K_\theta$), the lower band evolves toward zero energy, eventually crossing it.*[69]

Figure 7h shows the local density of states of the flat bands, where clear changes in amplitude exist between the different stacking domains (e.g., AA, AB, BA). Figure 7i is a diagram showing the different stacking domains. Topological order forms across the different domains, including 1D conducting channels, which is discussed more later.

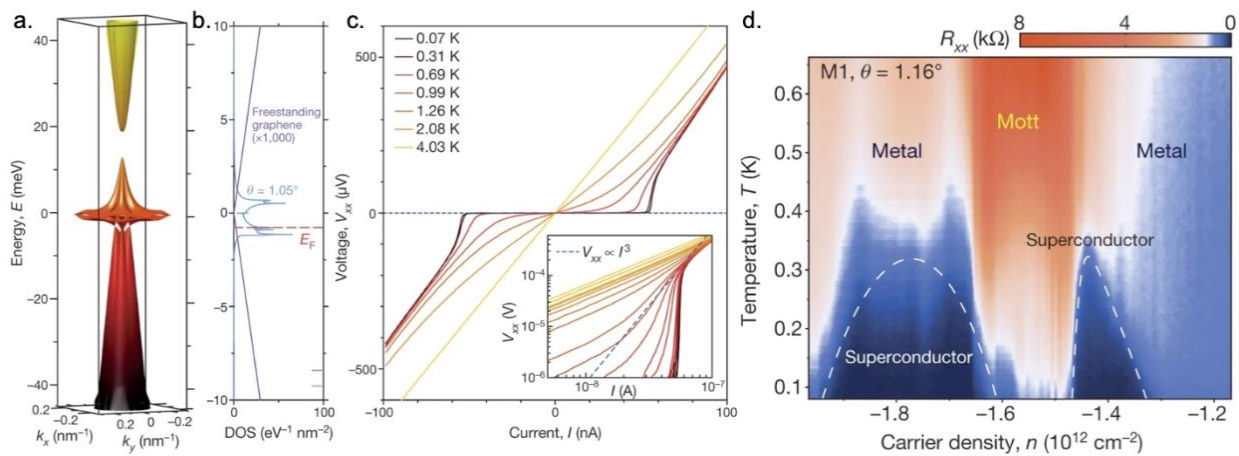

**Figure 8: Gate-tunable superconductivity at "magic" angles in tBLG.** (a) The band energy $E$ of tBLG at $\theta = 1.05°$ in the first mini Brillouin zone of the superlattice. The bands near charge neutrality ($E = 0$) have energies of less than 15 meV. (b) The density of states (DOS) corresponding to the bands shown in (a) for energies of −10 to +10 meV (blue; $\theta = 1.05°$). For comparison, the purple lines show the total DOS of two sheets of freestanding graphene without interlayer

interaction (multiplied by 10³). The red dashed line shows the Fermi energy $E_F$ at half-filling of the lower branch ($E<0$) of the flat bands, which corresponds to a density of $n = -n_s/2$, where $n_s$ is the moiré superlattice density (defined in the text). Superconductivity is observed near this half-filled state. (c) Current–voltage ($V_{xx}$–I) curves for a device measured at $n = -1.44 \times 10^{12}$ cm$^{-2}$ and various temperatures. At the lowest temperature of 70 mK, the curves indicate a critical current of approximately 50 nA. The inset shows the same data on a logarithmic scale, which is typically used to extract the Berezinskii–Kosterlitz–Thouless transition temperature ($T_{BKT} = 1.0$ K), by fitting to a $V_{xx} \propto I^3$ power law (blue dashed line). (d) Four-probe resistance $R_{xx}$ of carrier densities vs. temperature. Two superconducting domes are observed next to the half-filling state, which is labelled 'Mott' and centered around $-n_s/2 = -1.58 \times 10^{12}$ cm$^{-2}$. The remaining regions in the diagram are labelled as 'metal' owing to the metallic temperature dependence. The highest critical temperature observed in the corresponding device is $T_c = 0.5$ K (at 50% of the normal-state resistance). Reprinted images and portions of the caption with permission from [70].

Figure 8a shows a portion of the theory calculated band structure of tBLG at the magic angle 1.05°, where the flat bands at zero Fermi energy enable near-zero velocity electrons across all directions. Figure 8b shows the density of states (DOS) near the charge neutrality point in Figure 8a. Superconductivity emerges when the Fermi energy ($E_F$) is brought to near half-filling of the lower flat band, as shown in Figure 8b. No appreciable superconductivity was observed when the Fermi energy was tuned into the upper flat band (i.e., the conduction band). Of note, the DOS increases more than ×1000 in tBLG with a strong vs. weak interlayer coupling, likely due to a Fermi velocity reduction and increase in localization that occurs near the magic angle. Figure 8c show the current-voltage measurements at different temperatures, exhibiting the typical behavior of a 2D superconductor. The measurements are in agreement with Berezinskii–Kosterlitz–Thouless (BKT) 2D superconductivity,[157,158] and show a transition temperature of $T_{BKT} \approx 1.0$ K at $n = -1.44 \times 10^{12}$ cm$^{-2}$. Figure 6d shows the four-probe resistance of tBLG at 1.16° as a function of carrier density and temperature, where two pronounced superconducting domes on each side of the half-filling correlated insulating state are present, a feature associated with high-temperature superconductivity in cuprite materials.

Here we define $n_s = 4/A$ to be the density that corresponds to full-filling of each set of degenerate superlattice bands, where $A$ is the area of the moiré unit cell, a = 0.246 nm is the lattice constant of the underlying graphene lattice and θ is the twist angle.

Theory predicts twist-dependent band flattening and superconductivity in a variety of vertically-stacked 2D structures, including bilayer TMDs,[159,160] trilayer graphene-boron,[161] and bilayer hBN.[162] Additionally, twist-dependent topological properties have been predicted in twisted TMD homobilayers.[163] To the best of the authors' knowledge, as of publication, band flattening has not been experimentally demonstrated in 2D materials outside of tBLG. However, previous work found signatures of superconductivity in twisted bilayer $WSe_2$.[164,165] This is in part due to the difficulty of such measurements and sample fabrication. Previous experimental work demonstrated that changes in twist angle as sensitive as 0.1° induced large changes.[67,69,70,98]

### 3.4.1. Strong electron correlations probed using optical methods in TMDs

Previous work demonstrated that strongly correlated electrons in 2D materials, due to the twist angle and moiré pattern, can evoke incredible new properties. This has been largely demonstrated using sensitive transport measurements; however, not all materials are conducive to such setups, suggesting new methods are required to probe the capabilities. An optical method has been developed to probe the strong correlations and interaction-induced incompressible states of electrons in TMD-based 2D heterostructures.[166] In $MoSe_2$/hBN/$MoSe_2$ 2D heterostructures twisted to an angle close to 0°, this method demonstrated strong layer pseudospin paramagnetism, and an incompressible Mott-like state of electrons, suggesting it can be used to probe Bose-Fermi mixtures of degenerate electrons and dipolar excitons.[166]

### 3.5. Twist-dependent transport: one-dimensional topological channels

Topological order is an effect that can induce exotic and new phases of matter in materials.[167] The long-range entanglement of quantum matter holds promise for transformative technologies such as quantum computing with decreased decoherence.[168] One-dimensional (1D) topological channels are of interest because they could act as "wires" where the topological effects are purely confined to a narrow region, facilitating device fabrication. Although 1D topological channels have been demonstrated in other materials,[169,170] to the best of the authors' knowledge, tBLG is the first and only experimental demonstration in 2D materials.[151]

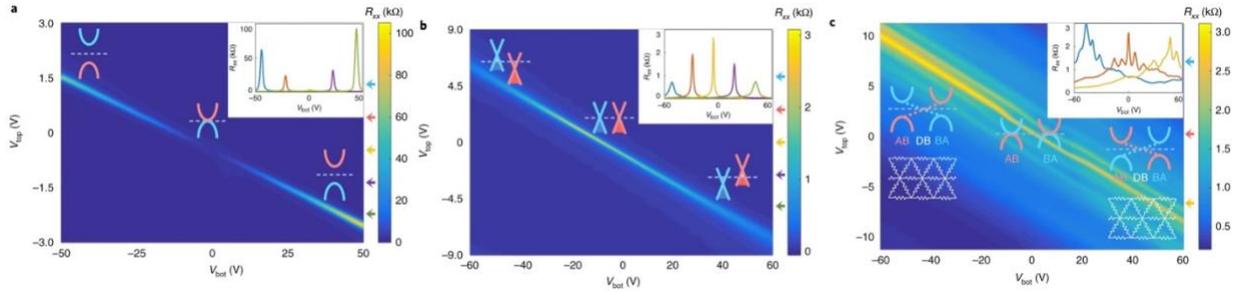

**Figure 9: Electronic reconstruction and 1D topological channels in tBLG, due to atomic reconstruction.** (a)-(c) Top- and bottom-gate dependence of the longitudinal resistance $R_{xx}$ in three different tBLG configurations: (a) Bernal-stacked, (b) large twist angle (2.8°), and (c) and small twist angle (0.47°). The insets show several line cuts at fixed top-gate voltages marked with corresponding colored arrows on the right side of the main panels. Schematic band structures in the inset show how the band structures change as a function of the perpendicular electric field. The schematic diagram of the triangular network of resistors in (c) represents the current paths generated along the domain boundaries (DB) obtained by gapping out AB and BA domains with transverse electric field. The transport along tBLG moiré grain boundaries has been experimentally demonstrated to be topological,[66,151] and theoretically predicted.[171] Reprinted images and portions of the caption from [98].

Figure 9 contains reprinted work demonstrating the significant effect atomic reconstruction has on the electronic properties of tBLG, and how the electronic properties evolve with twist angle.[98] Massive changes are observed at the 0.47° twist angle, compared to the 0° and 2.8° twist angles, likely due to the strong correlations. A subtle and important finding was the discovery of the topological insulating states that exist along the domain boundaries in tBLG near magic twist angles, thereby producing one-dimensional (1D) topological insulating channels.[66,98,151,171] Figure 9c shows the resistor network - likely from 1D topological channels - that forms to produce the transport measurements.

### 3.6. Moiré superlattice exciton quasiparticles in bilayer TMD 2D structures

As discussed in the introduction, a variety of monolayer TMDs exhibit a bright photoluminescence (PL) due to tightly bound exciton quasiparticles, and a direct band gap.[172,173] When monolayer TMDs are vertically-stacked, and the interlayer coupling is made sufficiently strong, twist-dependent moiré superlattice exciton quasiparticles can emerge. This effect has been demonstrated experimentally in a variety of twisted bilayer TMD 2D structures,[80–83] and is agreement with robust theory works.[84,85]

As the two TMD layers form a sufficiently strong coupling, interlayer hybridization fosters the formation of interlayer excitons that reside in the interlayer region. More specifically, the electron in one layer forms a bound pair with the hole in the other layer.[84] Theory makes a strong case that excitons only form when the electron and hole reside at the same momentum, suggesting stable excitons only exist across direct bandgaps. Despite this, published works report on interlayer excitons that form across an indirect bandgap in the interlayer region;[81,82,174–176] photoluminescence peaks are observed that correspond to a theory-predicted indirect transitions. It should be noted that it is very unlikely that the indirect exciton is truly indirect because this would require a significant shift in momenta, and a mechanism for this has not be identified. To overcome this, unique selection rules for interlayer excitons have been proposed, where momenta is fully conserved.[177]

The interlayer hybridization, and formation of direct bandgaps, is dependent on the atomic registry (*i.e.,* how the atoms between the layers overlap), which is depend on the ratio of lattice constants and twist angle. This effect is similar, but not identical, to the band flattening observed in tBLG. In tBLG, an entirely new band structure forms that overtakes the band structure of either graphene layer individually. In contrast, the band structures and properties corresponding exclusively to the individual TMD parent layers can still be detected, alongside interlayer and moiré superlattice effects that are the result of the interaction between the layers. When the interlayer coupling is mitigated, the interlayer effects disappear, leaving only a simple summation of the parent properties.[72,91,178]

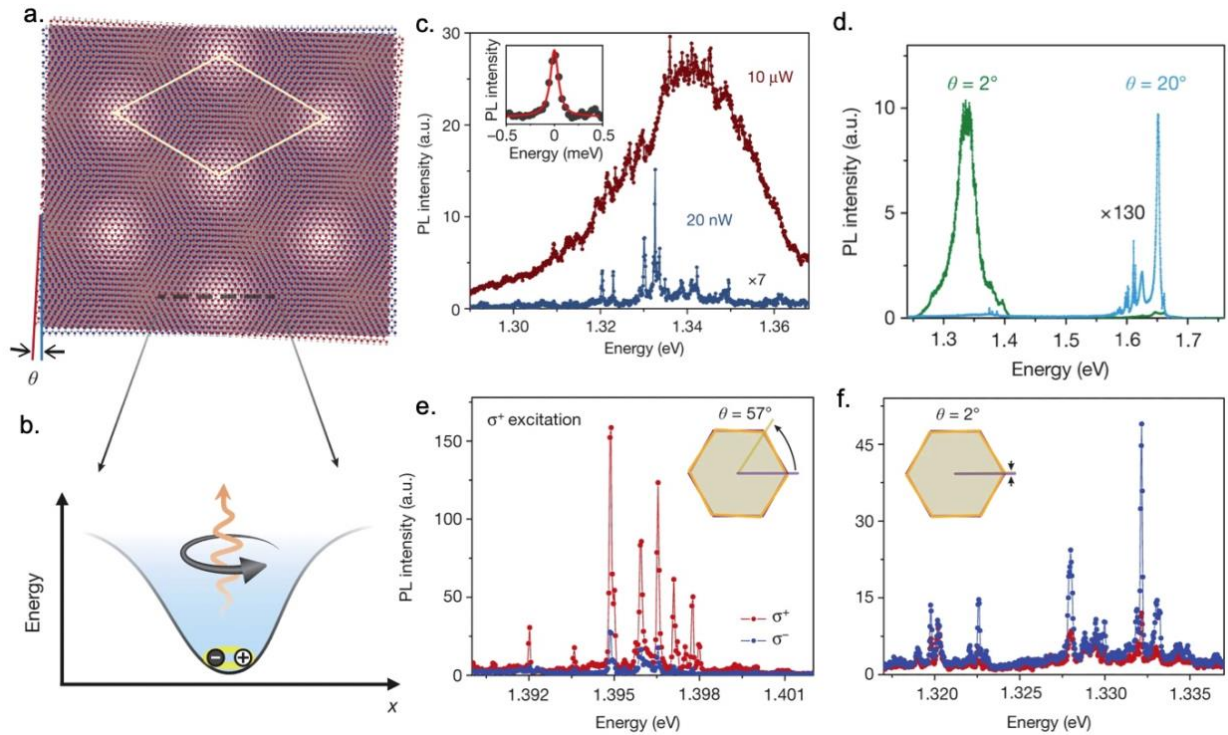

**Figure 10: Moiré excitons in twisted bilayer TMD 2D structures.** (a) Diagram of a representative moiré pattern formed in a twisted bilayer TMD 2D heterostructure with twist angle θ. (b) Schematic of an exciton quasiparticle trapped in a moiré superlattice potential site. (c) Interlayer exciton photoluminescence (PL) from a bilayer $MoSe_2$/$WSe_2$ 2D heterostructure with a 2° twist angle illuminated at two different powers: 10 μW (dark red) and 20 nW (blue; intensity scaled by ×7). Inset, Lorentzian fit to a representative photoluminescence peak suggests a linewidth of ~100 μeV. (d) PL spectra from a bilayer $MoSe_2$/$WSe_2$ 2D heterostructure with twist angles of approximately 2° (green) and 20° (blue; intensity scaled by ×130), at an excitation power of 5 μW. (e)-(f) Helicity-resolved PL spectra of bilayer $MoSe_2$/$WSe_2$ 2D heterostructures twisted to (e) 57° and (f) 2°. The trapped interlayer excitons are probed by illuminating the sample with σ⁺-polarized light at 1.72 eV. The σ⁺ and σ⁻ components are shown in red and blue, respectively. The PL at 57° is co-circularly polarized, while 2° is cross-circularly polarized. The changes are in general agreement with DFT calculations showing strong valley polarization in monolayer TMD. Reprinted images and portions of the caption from [82].

Although most excitons are the bound pair of one electron and one hole, there are other types of exciton quasiparticles. For example, trions are three particles bound together, where a positive trion is two holes and one electron, while a negative trion is two electrons and a hole. Additionally, excitons can break spin-symmetry by being polarized or spin-dependent.[179]

With respect to moiré excitons, the moiré superlattice determines the type of interlayer exciton formed. The interlayer atomic registry, long-range moiré periodicity, and strength of the interlayer coupling all affect interlayer excitons.[80–85] Scanning tunneling microscopy (STM) measurements show a periodically varying interlayer separation and electronic bandgap that follows the moiré superlattice formed by bilayer MoS$_2$/WSe$_2$ 2D heterostructures.[143] This leads to the following statement:

*The new long-range periodicity facilitates the formation of moiré minima, or moiré wells, that induce confinement.*

Figure 10a shows a representative moiré pattern, and Figure 10b is a cartoon depicting the a moiré minima with a moiré exciton confined inside. Previous work suggests that moiré wells maintain three-fold rotational (C3) symmetry,[84,180] indicating moiré-trapped interlayer excitons inherit valley-contrasting properties,[82] a property that sharply distinguishes it from general excitons bound to randomly formed extrinsic potential traps. More specifically, it has been shown that excitons can be trapped in different potentials; however, a key difference is that a moiré-trapped exciton inherits properties of the moiré superlattice.

The unique nature of moiré minima is demonstrated in Figures 10c-10f. Figure 10c demonstrates how the broad interlayer PL peak decomposes into multiple narrow peaks near the free interlayer exciton energy (~1.33 eV), that have a line-width comparable quantum emitters fabricated using 2D materials.[34,35,181]

Figure 10d shows PL spectra from two different twist angles - 2° and 20° - where significant changes to the moiré exciton are identified. Although the characteristic monolayer PL peaks of both parent materials are quenched – likely due to ultrafast interlayer charge transfer[182] – the moiré exciton PL peak is present. Of note, the moiré exciton is more than two orders magnitude brighter when twisted to 2° vs. 20°, likely due to changes in how the electron and hole Brillouin zones overlap. More specifically, each parent layer contributes either a hole or an electron, and a Brillouin zone for each layer. When the corners match closely – enabling the formation of interlayer high symmetry points – the hole-electron pairs are close in real space, and require only a minimal momentum shift in momentum space.[180] This happens at twist angles close to 0°. Conversely, when twisted to 20°, the corner do not overlap and a sufficiently dense number of high symmetry points are not formed, leading to a diminished PL intensity.

Using circular-polarization-resolved PL, Figure 10e and 108-8f demonstrate how trapped interlayer moiré excitons exhibit strong valley polarization. Twist angles near 57° display over 70% valley polarization due to selection rules based on how the Brillouin zones overlap.[82] When twisted to near 2°, the selection rules are reversed, as demonstrated in Figure 10f. The moiré exciton circular polarization response is different from other 2D quantum emitters, possibly due to the anisotropy of the interlayer trapping potential, which breaks the three-fold ($C_3$) rotational symmetry of the host lattice.[183] Such work holds promise for twistronics technology because it suggests the properties of the trapped moiré excitons are determined by the twist-dependent local interlayer atomic registry.

### 3.7. Twist-dependent ferromagnetism in tBLG

One of the more surprising discoveries is the observation of emergent ferromagnetism in a unique 2D heterostructure: tBLG twisted to a magic angle near 1.15°, and then monolayer hBN twisted to align with the top graphene layer is placed on top (i.e., twist angle = 0°).[67,86] The interlayer coupling induced strong electron correlations likely facilitate orbital magnetism, due in part to a robust easy-axis anisotropy arising from the 2D nature of the graphene bands.[67,68,87] Of note, the ferromagnetism only emerges when the 2D structure is gated to a ¾ filling of the conduction band, which corresponds to an insulating state. The effect only presents itself when all three layers are twisted to the correct angles, suggesting the field of twistronics will grow beyond simple twisted bilayer structures into multilayer structures with unique twist angles for each layer.

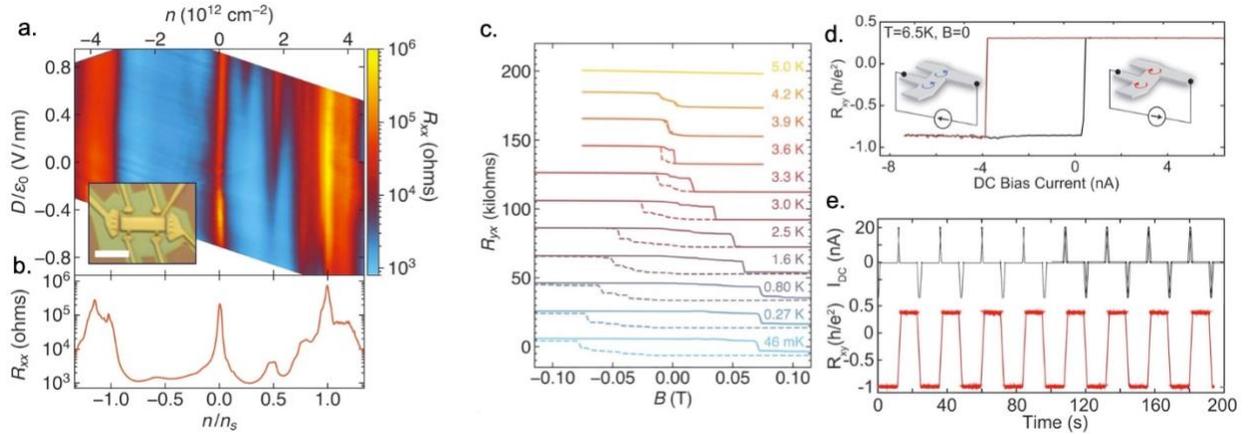

**Figure 11: Magic-angle-dependent correlated states facilitate emergent ferromagnetism, and current-controlled magnetic switching.** (a) Longitudinal resistance $R_{xx}$ of tBLG at 2.1 K as a function of carrier density $n$ and perpendicular displacement field $D$, tuned using top- and back-gate voltages. $n$ is mapped to a filling factor relative to the moiré superlattice density $n_s$, corresponding to four electrons per moiré superlattice unit cell, shown on the bottom axis. (Inset) Optical micrograph of the completed device showing the top-gated Hall bar. Scale bar, 5 μm. (b) Line cut of $R_{xx}$ with respect to $n$ taken at $D/\epsilon_0 = -0.22$ V/nm showing the resistance peaks at full filling of the superlattice and additional peaks likely corresponding to correlated states emerging at intermediate fillings. (c) Temperature dependence of $R_{yx}$ versus $B$ at $D/\epsilon_0 = -0.62$ V/nm and $n/n_s = 0.746$, showing the hysteresis loop decreasing as the temperature is raised. Successive curves are offset vertically for clarity. (d) $R_{xy}$ as a function of applied direct current, showing hysteresis as a function of direct current analogous to response to an applied magnetic field. Insets show schematic illustrations of current-controlled orbital magnetism. (e) Nonvolatile electrical writing and reading of a magnetic bit at T = 6.5 K and B = 0. A succession of 20-nA current pulses of alternating signs controllably reverses the magnetization, which is read out using the Hall resistance. The magnetization state of the bit is stable for at least $10^3$ s. Reprinted (a)-(c) and portions of the caption from [67], and reprinted (d)-(e) and portions of the caption from [86].

Figure 11a shows the longitudinal resistance of a tBLG/hBN 2D structure as a function of carrier density $n$. The carrier density was tuned using top and bottom gates. Note the insulating states are certain fractional fillings of the conduction band, quantified using the filling factor $n/n_s$, where $n_s$ is the moiré superlattice density of four electrons per moiré superlattice unit cell. Figure 11b is a line cut at $D/\epsilon_0 = -0.22$ V/nm, demonstrating the carrier density-dependent insulating states due to the strong electron correlations. Figure 11c shows the effect of an external magnetic displacement field on the tBLG longitudinal resistance, where characteristic hysteresis loops are observed. Sharpe *et al* found that the hysteresis loop only emerges in a narrow range of densities near ¾ filling, despite the fact that it is observable over a wide range of displacement fields. Together, this is unambiguous evidence of ferromagnetism.

The ferromagnetic domain polarization can even be controlled using very small direct currents (~4 nA) because the tBLG domains interact with applied current. Figure 11d shows the resistance switching based on the applied current, where a sharp hysteretic switching between magnetization states is observed. Once switched, the state appears to be indefinitely stable.[86] Figure 11e shows deterministic writing and reading of magnetic bits using current pulses and the resulting change in the anomalous Hall resistance. Exceptionally small currents are used to write the magnetic bits (~20nA), while even small currents are required to read (<100pA), demonstrating exceptional promise for ultralow power computing. The pico-amp read currents are particularly exciting because reads are generally far more frequent than writes.

### 3.8. Twist on Raman and phonon modes

Raman spectroscopy, or Raman scattering, measures the phonon and vibrational modes in a material, which are dependent on the distribution of atoms and charge.[184] It is a powerful technique to evaluate 2D materials because it is non-destructive, non-intrusive, and facile, and can probe numerous properties,[185] including the in-plane symmetry,[186] defects,[187] strain,[188,189] doping,[189,190] interlayer coupling,[76] charge interactions,[191] charge density waves,[192] and the twist-angle.[93–96,135,172]

Not only can polarized Raman spectroscopy measure the twist angle directly,[193] but it can identify new Raman modes that manifest purely due to the interlayer coupling and twist-angle.[55,61–63,172] The interlayer coupling redistributes charge into the interlayer region (Section 3.2),[55] modifies electronic bonding, and encourages the reconstruction of the moiré superlattice (section 3.3),[98,144] which, together, can induce detectable changes in the Raman modes.

## 4. Twisted 2D structure fabrication methods

This chapter strives to be a very comprehensive summary of demonstrated methods to fabricate twisted 2D structures. This includes not only growth and mechanical stacking methods, but also techniques on twisting the 2D structures to new angles.

### 4.1. Mechanical transfer by vertically-stacking one material on another

One of the simplest, most successful, and most popular methods for fabricating 2D structures is a simple mechanical transfer where the 2D materials are vertically-stacked on another using a physical transfer process. As a testament to the methods effectiveness and success, nearly all of the top discoveries in the field of twistronics applied this method, including superconductivity,[69,70] orbital magnetism,[67] Hofstadter's butterfly,[97] and atomic reconstruction.[98,144] The method has also demonstrated success with certain magnetics materials that are more sensitive to foreign contaminants in ambient, enabling a variety of high-profile spintronics applications, including giant tunneling magnetoresistance,[16] and switching magnetic states through pressure tunning.[75]

The method's biggest weaknesses is that it is not scalable, and that it requires expensive equipment and knowledge infrastructure. Researchers have mitigated these concerns by developing machine-learning assisted robots that are able to exfoliate and characterize samples, and then assemble them into user-specified 2D structures.[194] Additionally, micro-stamper technology demonstrated "2D material printing" of mechanically stacked structures with promise for industrial prototyping applications.[195] Despite these impressive results, likely more work needs to be done to make the mechanical transfer method scalable and cost-effective beyond single and niche research applications.

### 4.1.1. Deterministic dry transfer using viscoelastic stamping

Monolayer materials are highly sensitive to their environment, making them ideal candidates for next-gen sensors. They have demonstrated a diverse multitude of chemical sensing applications,[20] including water vapor,[37] nerve agents,[21] oxygen,[61] toxic gases (e.g., $NO_2$ and $NH_3$),[23] and individual ions.[22] Unfortunately, these same attributes that enable advanced sensing, make them highly-sensitive to and easily functionalized from chemicals in their environment, suggesting wet mechanical transfers (i.e., those where a liquid contacts the 2D material) likely alter or destroy the pristine nature of the parent 2D materials.

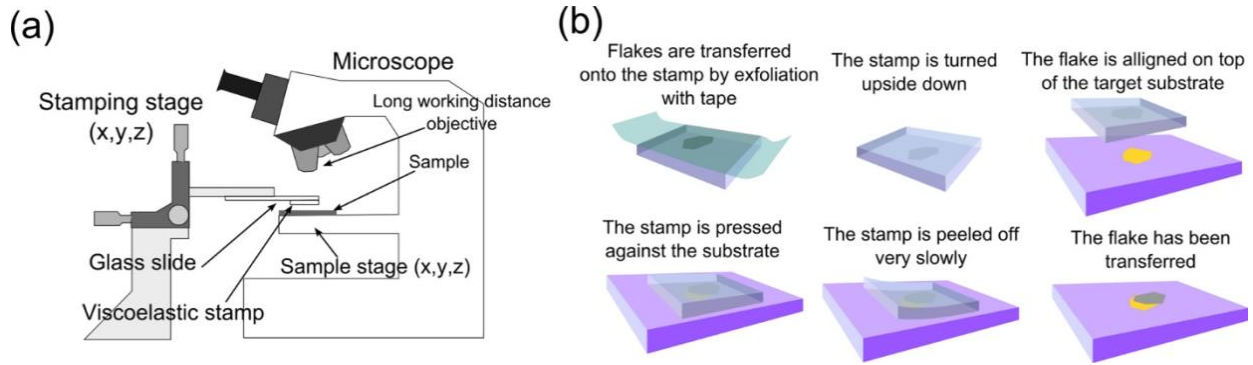

**Figure 12: Deterministic dry transfer process to mechanically stack 2D materials using a viscoelastic stamp.** (a) Schematic of the setup. (b) Diagrams showing the steps involved for deterministic transfer of an atomically thin flake onto a user-defined location, including another 2D material. Reprinted images and portions of the caption from [196].

To overcome this weakness, a dry transfer technique was developed using a viscoelastic stamp that does not use any wet chemistry and preserves the pristine properties of the transferred materials.[196] Previous work demonstrated transfer printing of non-2D materials using an elastomer stamp, where the stamp behaves as an elastic solid over short timescales, but can flow slowly over long timescales.[197] The difference in behavior breaks the surface adhesion symmetry allowing it to pick-up in one state (i.e., the elastic solid) and deposit in the other (i.e., while slowly flowing). More specifically, by slowly peeling off the stamp from the surface, the viscoelastic material detaches, releasing the flakes that adhere preferentially to the acceptor surface. It was found that the whole transfer can be accomplished in less than 15 minutes and largely appears to preserve the pristine properties of the monolayer materials.

### 4.1.2. High-accuracy rotational alignment

The previous section described how 2D materials can be mechanically stacked using a dry transfer method, but the twist-angle cannot be controlled in that setup. In fact, the exact crystal orientation is frequently not known and needs to be assumed using straight edges as a crystallographic reference. A limitation of this method is that the edges are not always straight, or purely one type of configuration. For example, graphene edges can be a mixture of zigzag and armchair. Additionally, optical resolution limits the accuracy of angle identification to about 2°.[198]

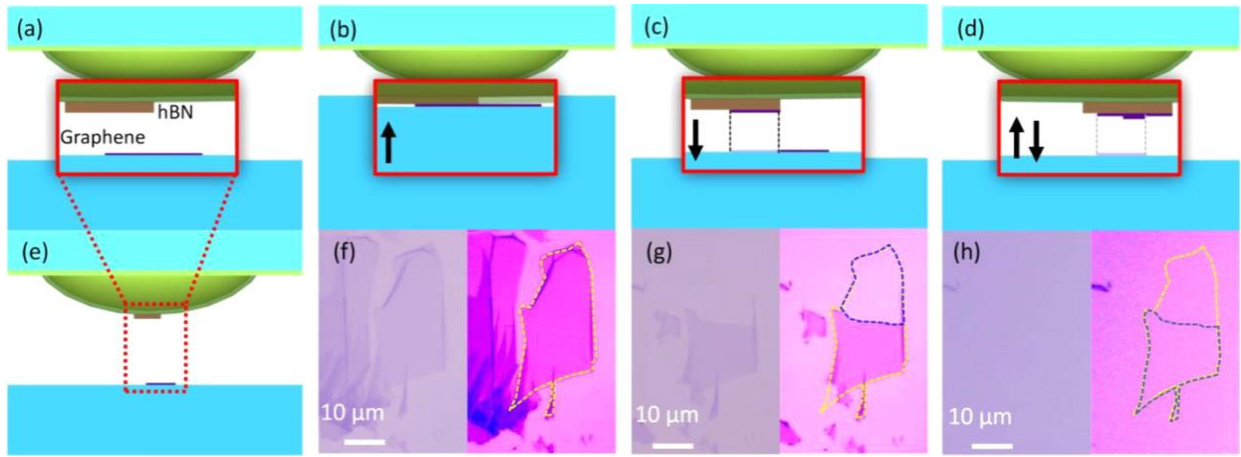

**Figure 13: High-accuracy rotational alignment enables user-determined twist-angles.** (a)-(e) Schematic of dry transfer layer pick-up using the hemispherical handle substrate method. Inset shows a zoom-in. (b)-(d) Schematics and (f)-(h) corresponding optical images of successive stacking steps with the contrast enhanced in the right image. (b) and (f) illustrate a partial contact of the handle with the bottom graphene. (c) and (g) show the handle substrate release with one graphene section detached. (d) and (h) illustrate the second contact with the handle translated laterally to create an overlap region of the two graphene layers. Because the two graphene layers are obtained from a single grain, their principal crystal axes remain aligned as long as the handle is not rotated with respect to the bottom substrate. If the handle is rotated, the user knows the exact twist (within a certain error). Reprinted images and portions of the caption from [198].

To overcome these challenges, a method was developed that tears a single crystal flake in half, and stacks the two pieces on top of each other (see Figure 13).[198] This method removes the requirement to measure and know the crystals' orientations in absolute space, taking advantage of the fact: deterministic control of the twist angle (i.e., the relative angle) is the goal. This method enables twist angles at zero degrees to be known with high confidence, and enables the ability to twist to very low angles with higher success. The method applies a dry method, thereby reducing the incorporation of residues or chemicals between the layers, in the interface.

### 4.1.3. AFM Nano-"Squeegee": Deterministically cleaning surfaces and interfaces

The presence of atoms or molecules in the interlayer region mitigates the interlayer coupling strength. The effect is very pronounced when contaminants or residue from the mechanical transfer process are trapped between the layers. A method was developed to clean the interfaces using an atomic force microscope (AFM), which presses down onto the top layer with sufficient strength to flatten it. The residue is then forced out similar to a squeegee.

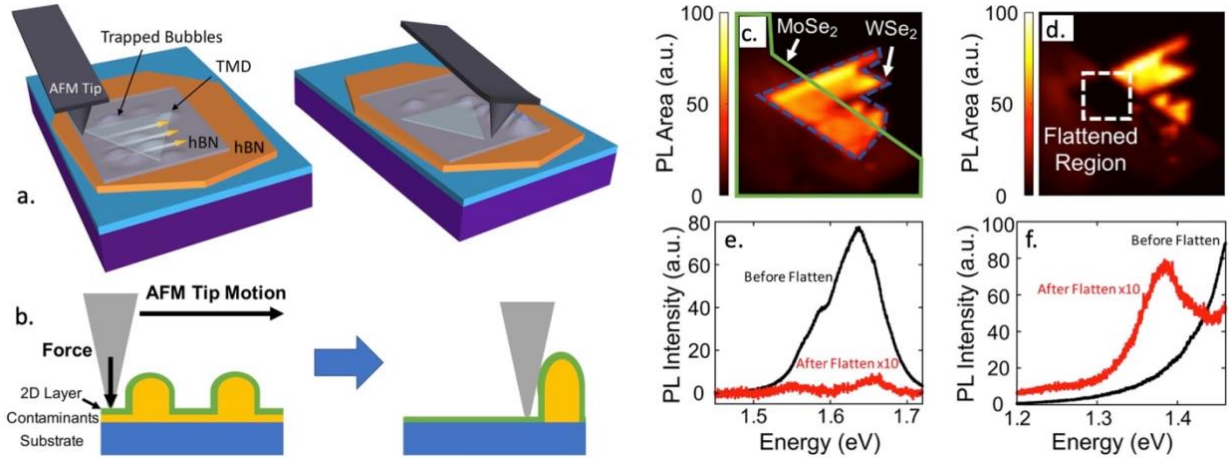

**Figure 14: Clean interfaces using AFM to nano-"squeegee" residue and contaminants away.** (a)-(b) Cartoons illustrating the general concept of nano-"squeegee" using an AFM. After mechanical transfer, there are contaminants trapped between the 2D layer and the substrate, which are often a combination of a uniform thickness layer and bubbles. Applying a normal load with the AFM tip pushes the contaminants out from beneath the tip and induces direct contact between the 2D layer and the substrate. Scanning the tip across the surface while applying normal load causes the contaminants to collect into a pocket, leaving behind a region without contaminants. (c)-(f) A 1L $MoSe_2$ + 1L $WSe_2$ 2D heterostructure on multilayer hBN (c),(e) before and (d),(f) after nano-"squeegee" in a select area. Spatial map of photoluminescence integrated area from 1.45 to 1.7 eV before flattening (c) and after flattening (d). (e) Characteristic PL spectra of $MoSe_2$ and $WSe_2$ intralayer excitons in the overlapping region before and after flattening. (f) Spectrum of $MoSe_2$/$WSe_2$ 2D heterostructure before and after flattening showing a feature at 1.38 eV, which corresponds to the energy of interlayer excitons. Reprinted images and portions of the caption from [72].

Figures 14a-b show cartoons demonstrating the nano-"squeegee" technique used to reproducibly and deterministically clean interfaces with high spatial control. Figures 14c-d show photoluminescence (PL) mapping of a 1 layer $MoSe_2$ + 1 layer $WSe_2$ 2D heterostructure before and after nano-"Squeegee", where the affected area shows a noticeable decrease in intralayer excitons from the monolayer $WSe_2$. Figure 14e show representative PL spectra from before and after nano-"squeegee", where the intralayer excitons corresponding to the monolayer configuration of $MoSe_2$ and $WSe_2$ are quenched after nano-"squeegee". These results are in agreement with previous findings that show bilayer configurations have a quenched PL relative to the monolayer configurations. Figure 14f shows the emergence of an interlayer exciton (i.e., an exciton quasi particle that is bound to a hole and electron that reside in different layers) after nano-"squeegee", suggesting the interlayer coupling was significantly strengthened.

As discussed in Section 4.4.1, previous work demonstrates that the interlayer coupling can be controllably manipulated using the controlled intercalation of foreign atoms and molecules.[58,61]

These foreign compounds likely disrupt charge distribution in the interlayer region and possibly push the layers farther apart, thereby decreasing the interlayer coupling strength.

### 4.1.4. Deterministic, *in situ* twisting using an atomic force microscope (AFM)

Previous publications demonstrated that a monolayer material can be pushed into new twist angles, relative to an underlying, continuous substrate,[100,137] suggesting that this technique can be applied to twist the top layer of arbitrary 2D structures into user-defined configurations.

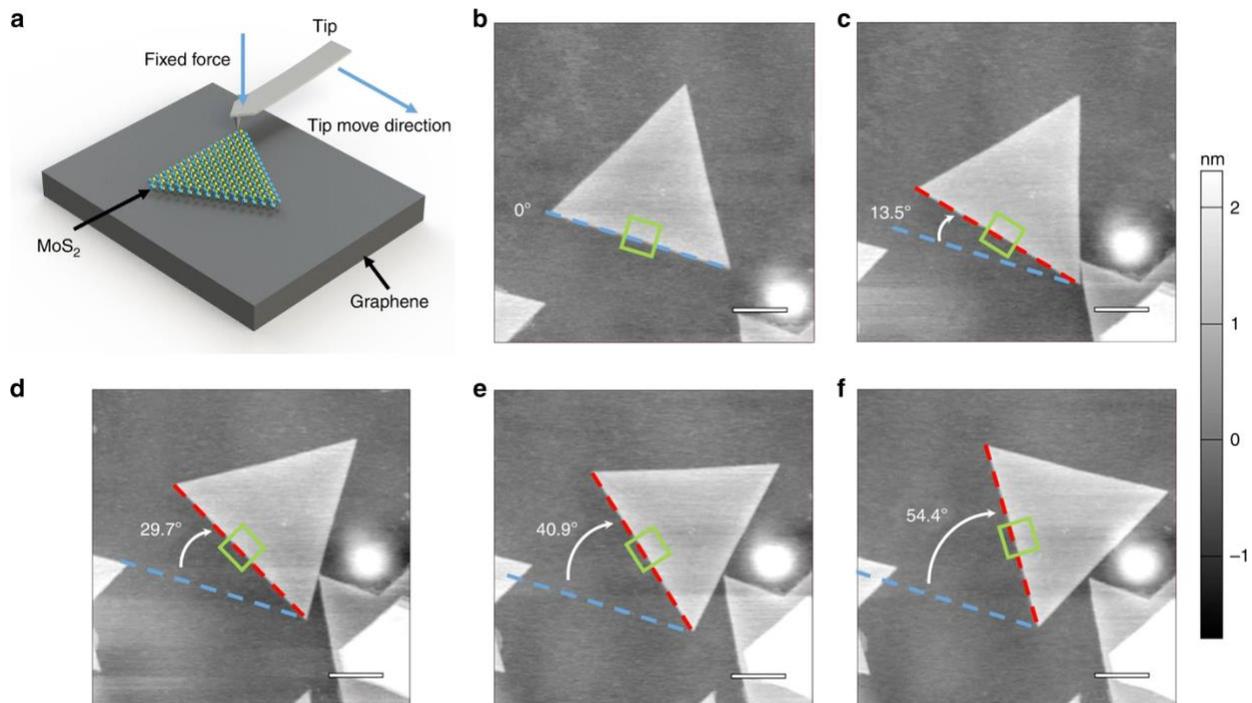

**Figure 15: Pushing and twisting monolayer materials using an AFM tip.** (a) Schematic demonstrating pushing a monolayer material into a new twist angle using an AFM-tip. (b)–(f) AFM images of a typical MoS$_2$ domain rotated on graphene to achieve a series of twist angles, scale bar, 1 μm. Blue dash lines indicate the original direction of the MoS$_2$ domain, while white arrows indicate the rotation directions. Reprinted images and portions of the caption from [100].

Figure 15 demonstrated a monolayer MoS$_2$ crystal being pushed and twisted using an AFM tip on graphene, thereby controllably manipulating the twist angle of a MoS$_2$/graphene 2D structure. When combined with the high-accuracy rotation alignment (Figure 13), users know the starting twist angle with high precision, thereby overcoming the issues associated with edge detection. The work found that the vertical conductivity of the MoS$_2$/graphene heterojunction can be tuned by approximately a factor of ×5 under different twist configurations, where the highest and lowest conductivities occur at twist angles of 0° and 30°, respectively. DFT calculations suggest that this conductivity change originates from the twist-dependent transmission coefficient through the interlayer region.

### 4.2. Growing twisted 2D structures

2D structures can be directly grown using chemical vapor deposition (CVD),[199] molecular beam epitaxy (MBE),[200,201] and atomic layer deposition (ALD),[202] where CVD appears to have the most publications and demonstrated the most diverse set of 2D heterostructures.[203–206] Growing twisted 2D structures is more scalable and more likely to produce a clean interface, compared to mechanical transfer, but suffers from a major hurdle:

> *The twist-angle cannot be controlled during the growth, but is a statistical process, where energetically preferred twist-angles are more likely. The energy of a configuration is determined by a confluence of factors, including the moiré superlattice, interlayer atomic registry, strain, and interlayer bond strength.*

The twist-dependent energy landscape contains energetically preferred configurations. In fact, the landscape can be sufficiently asymmetric, that 2D layers will even twist into more preferred configurations while in ambient or being thermally annealed.[62,99,205,207–209] To the best of our knowledge, no work has been published demonstrating a general, non-specific method to grow 2D structures at user-defined twist angles. Although the natural preference for certain twist angles can be beneficial, it clearly is also a challenge that impedes researches from exploring the full twist phase space. Despite these challenges, it has been shown that the twist-angle can be manipulated of as-grown 2D heterostructures using a focused electron beam (Section 4.3),[205] or an AFM (Section 4.1.4).[100,137]

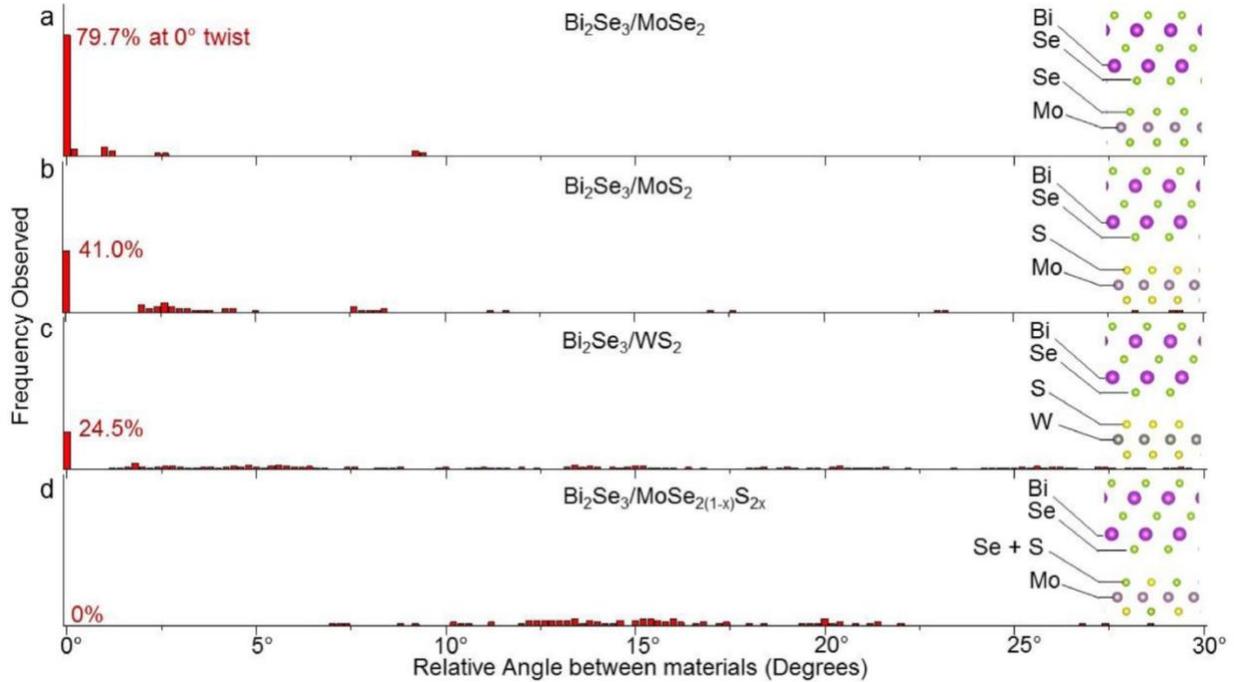

**Figure 16: Preferred twist angles (i.e., twist energy landscape) depends on the interlayer interaction.** Histograms of the twist angles ($Bi_2Se_3$ *vs.* the TMD) in as-grown 2D heterostructures (as labeled), with the following interlayer chemistry: (a) Se–Se interface with molybdenum. (b) Se–S interface with molybdenum. (c) Se–S interface with tungsten. (d) Se–Se/S–alloy interface with molybdenum. Despite their large crystallographic dissimilarities, $Bi_2Se_3$ prefers to grow near 0° when chemical periodicity is present in the underlying material (i.e., non-alloy heterostructures), but prefers 12°–20° when it is absent (i.e., $MoSe_{2-2x}S_{2x}$ alloy), suggesting chemical periodicity strengthens the interlayer interaction at 0°. The frequency at 0° decreases from (a)–(c) suggesting Se–Se interfaces and molybdenum-based TMDs form stronger interlayer interactions at this angle. Reprinted images and portions of the caption from [205].

Figure 16 shows histograms from different CVD-grown $Bi_2Se_3$/TMD 2D heterostructures, demonstrating differences in preferred twist angles.[205] The interfacial chemistry, geometric moiré patterns, and chemical periodicity (i.e., alloy *vs.* non-alloy) were all found to influence the interlayer interaction and twist energy landscape. Interestingly, for all non-alloy 2D structures, a twist angle of 0° is preferred, whereas no alloy 2D structures grew at 0°, suggesting chemical periodicity plays a significant role in the twist energy landscape.

### 4.3. Site-selectable, *in situ* twisting using a focused electron beam

Previous work demonstrated that sufficiently high electron beam currents can move and twist small crystallites on both amorphous[210] and crystalline substrates.[211,212] The mechanism is likely heating due to the electron beam. More specifically, at sufficiently high thermal energies, the particle overcomes the interaction and weak bonding with the substrate, enabling it to gain mobility and move across the substrate. The particle's mobility is dependent on the substrate's energy landscape, where, for example, certain twist angles are more stable than others.

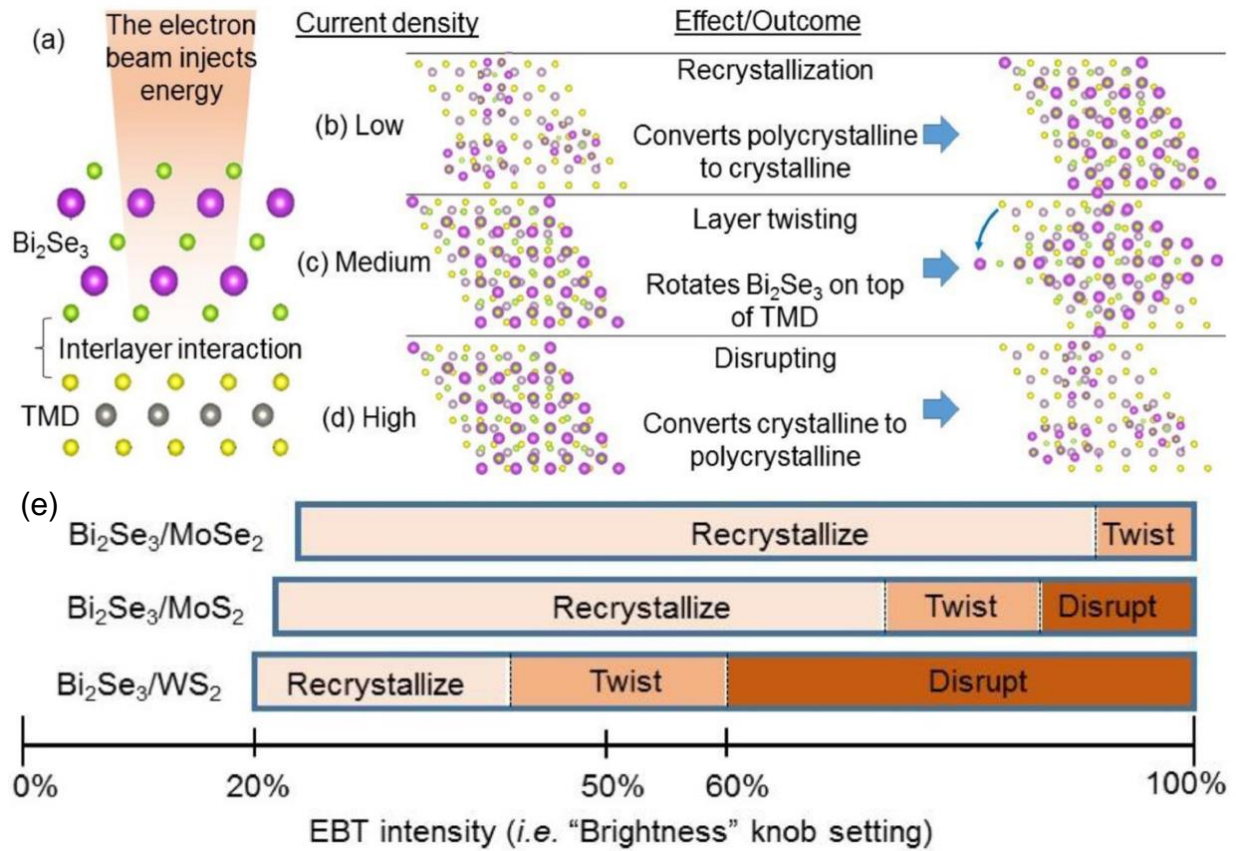

**Figure 17: Electron beam modifies the twist angle and structure.** (a) A vertically incident electron beam treatment (EBT) imparts thermal energy into a 2D heterostructure that—at sufficiently high current densities—induces it to overcome the interlayer interaction and rearrange. (b) Low current densities break only the weak interlayer bonds associated with semi-stable grains, resulting in recrystallizing (*i.e.*, grain migration to stable configurations). (c) Intermediate current densities overcome most interlayer bonding, but cannot disrupt the stronger intra-layer bonding, enabling significant grain rotation (twist). (d) Highest current densities break both the inter- and intra-layer bonding, disrupting the crystallographic order and leading to the formation of nano-crystals. (e) Summary of different 2D heterostructure's response to an EBT intensity. The twist angle can be modulated using the correct EBT intensity, as demonstrated in Figure 18. Reprinted images and portions of the caption from [205].

Previous work demonstrated that the twist angle can be controllably manipulated with high spatial selectivity in 2D materials using a focused electron beam (see Figure 17).[205] High-intensity electron beam treatments (EBT) break the intralayer bonds of one of the layers, facilitating the crystal to break-up into crystallites. Medium-intensity EBTs preserve the intralayer bonds, while breaking the weak interlayer bonds, enabling the one of the layers to gain mobility and twist across the other layer (see experimental data in Figure 18). Low-intensity EBTs gently perturb the layers into a stable twist angle.

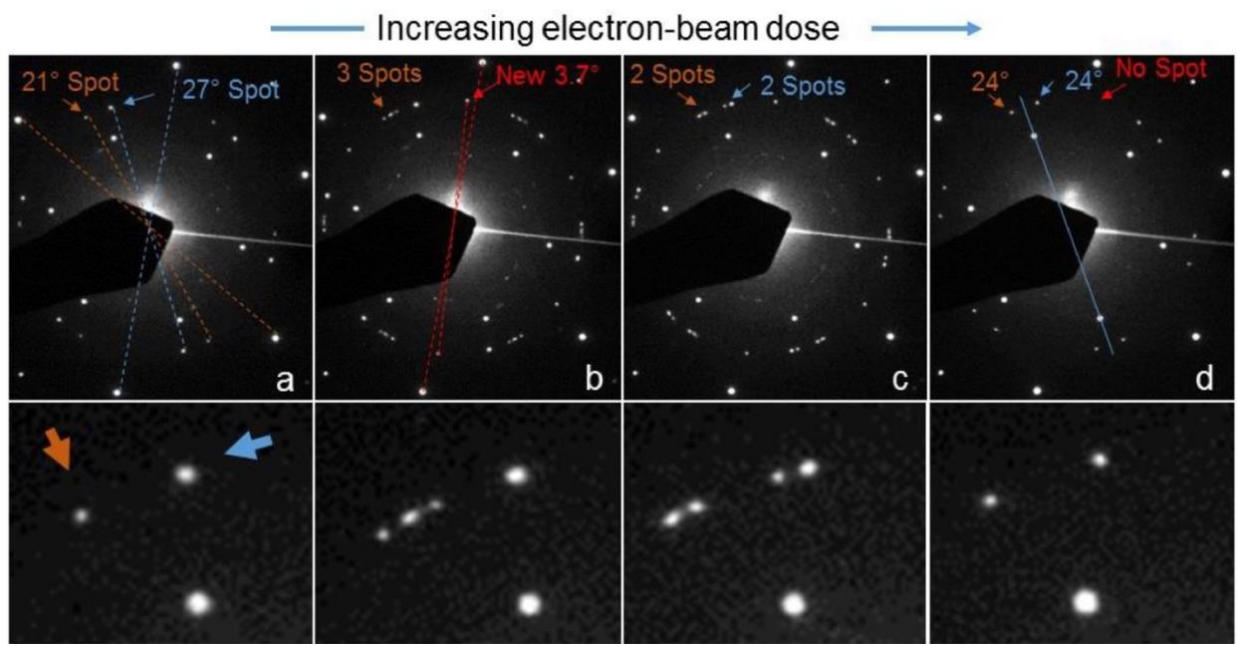

**Figure 18: Manipulating the twist angle *in situ* using an electron beam.** (a)–(d) TEM SAED images of a $Bi_2Se_3/MoS_2$ heterostructure subject to consecutive TEM electron beam treatments (EBTs). (a) Initially two $Bi_2Se_3$ spots are observed, where the spot at 27° is ×2.58 times brighter than the spots at 21°. (b) <u>After the 1st EBT:</u> The spot at 21° splits into 3 smaller spots, and a new spot appears at 3.7°. (c) <u>After the 2nd EBT:</u> The 3 spots near 21° combine into two, and the 27° spot splits into two. (d) <u>After the 3rd EBT:</u> Both sets of $Bi_2Se_3$ spots settle at 24°, and are nearly equally bright—8.4 versus 8.0 photon counts for orange and blue, respectively—suggesting the 24° twist angle is more stable than either 21° or 27. The 3.7° spot disappeared, suggesting the grain was perturbed out of the imaging area. Reprinted images and portions of the caption from [205].

Figure 18 shows experimental measurements demonstrating twisting of a $Bi_2Se_3/MoS_2$ 2D heterostructure using a focused electron beam in selected area electron diffraction (SAED) mode. Twisting of $Bi_2Se_3/MoSe_2$ and $Bi_2Se_3/WS_2$ 2D heterostructures have also been demonstrated.[205] A weakness of the method is that it cannot twist too large of areas because the electron beam currents required would likely destroy the 2D material. A strength of the method is that it can twist the materials *in situ* while remaining in the TEM, enabling a large number of twist angles to be probed rapidly, without having to remove the sample to affect a new twist angle. This method when coupled with dark field imaging[98,144], shows promise for rapidly collecting large amounts of data about atomic reconstruction and the interlayer interaction.

### 4.4. Tuning twist-dependent properties through interlayer coupling strength

Twist-dependent properties only emerge when the interlayer coupling is sufficiently strong. Tuning the interlayer coupling stronger or weaker deterministically switches the properties on or off. Two methods to tune the interlayer coupling are controlled intercalation, and hydrostatic pressure to force the layers together. Not only do they facilitate new technologies, but they are an experimental knob to tune the phase diagram of numerous capabilities, including photoluminescence from exciton quasiparticles,[61,63] superconductivity,[74] magnetism,[75,77] and Raman modes.[93–96]

#### 4.4.1. Controlled intercalation

Previous work demonstrated that intercalation of either atoms or molecules into the interlayer region, or between 2D materials and their substrate, disrupts the coupling and alters the properties. Below are three examples. First: when molecules are intercalated into the interlayer region of vertically-stacked vdW materials, the interlayer coupling is weakened, leading the individual layers to behave as if they were monolayer.[58] Second: When $O_2$ intercalates between a 2D material and their substrate, the 2D material decouples from the substrate and behaves more "free standing".[213–219] Third: Lithium forms itself into a superdense crystal when intercalated between bilayer graphene, disrupting the tBLG properties.[64]

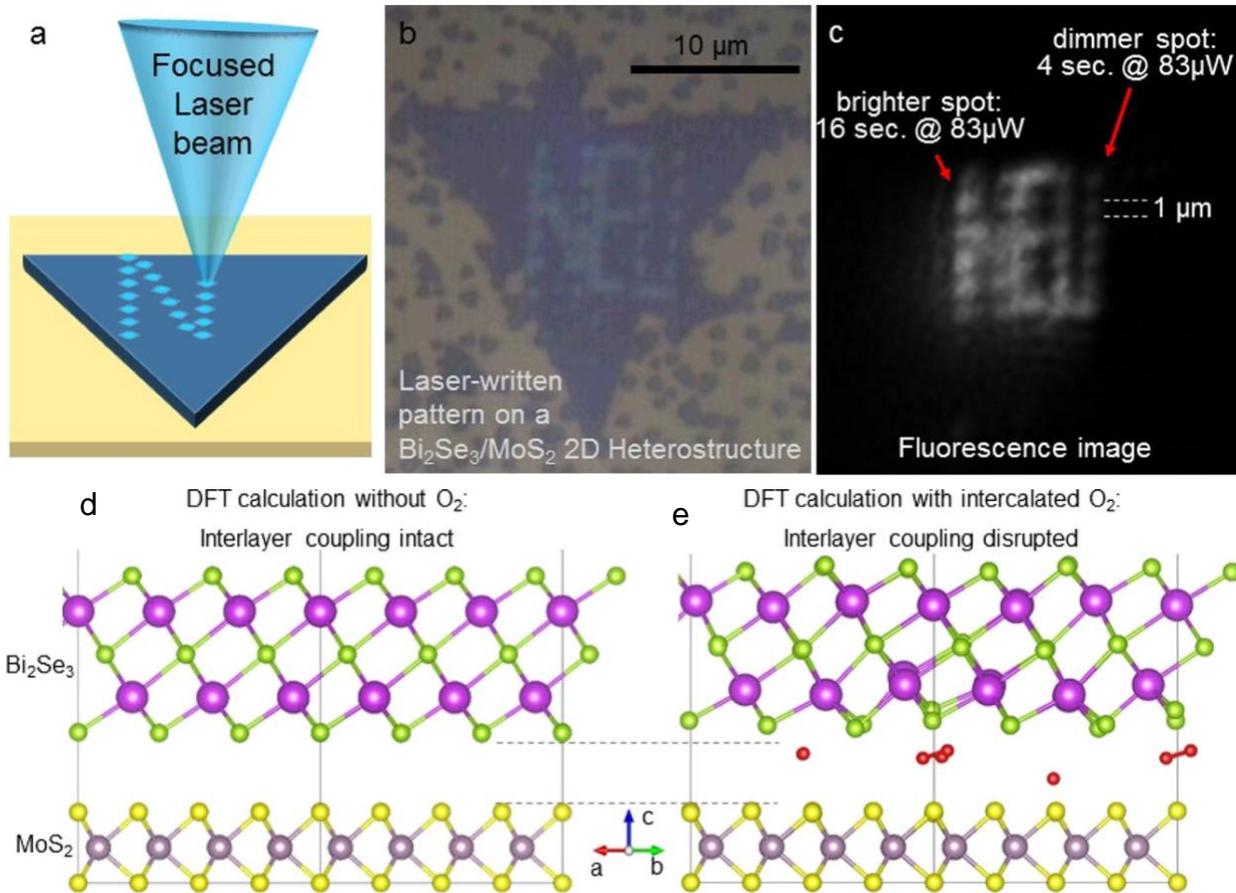

**Figure 19: Site-selective manipulation of interlayer coupling and photoluminescence through controlled intercalation of oxygen** (a) Schematic of a method to write patterns and tune the interlayer coupling strength on arbitrary $Bi_2Se_3$ + transition metal dichalcogenide (TMD) 2D heterostructures using a focused laser in ambient. (b) Optical image of a laser-written pattern on a $Bi_2Se_3$/$MoS_2$ 2D heterostructure. The letters "NEU" were "drawn" using different exposure times. (c) Fluorescence microscopy image of the same sample (excitation $\lambda$ = 488 nm). Note that using a focused laser beam, excitons of selected regions can be programmed to recombine radiatively (bright regions) or non-radiatively (dark regions), where the size of the affected area is dependent on the laser spot geometry and recipe used (*i.e.*, the power and exposure duration). The smallest "radiative" regions were below a micron in diameter. This method allows for the rapid manipulation and measurement of the interlayer coupling with high spatial resolution. (d)-(e) DFT calculations show intercalated $O_2$ diminishes the interlayer coupling. (d) DFT calculations of a rotationally aligned (*i.e.*, twist angle is 0°) $Bi_2Se_3$/$MoS_2$ superlattice predict significant charge redistribution into the interlayer region and an influential interlayer coupling (see Figure 4 or reference [55]). (b) When $O_2$ molecules are intercalated into the interlayer region, DFT calculations show that the average interlayer separation increases from 3.57 to 4.18 Å (17% increase), thereby diminishing the interlayer coupling. Amplifying information on other $Bi_2Se_3$/TMD 2D heterostructures can be found in [63]. Reprinted images and portions of the caption from [61].

$Bi_2Se_3$ has a natural affinity towards oxygen, where previous work demonstrated that it is able to easily diffuse into and through the material.[220] Figure 19 shows how the interlayer coupling of $Bi_2Se_3$/TMD 2D heterostructures can be deterministically modulated by controllably intercalating oxygen. The photoluminescence (PL) intensity and exciton recombination energy are used as metrics for the interlayer coupling strength, where the exciton recombination energy shifts with increasing oxygen intercalation.[63] For these experiments, the 2D heterostructures consisted of a monolayer TMD crystal with mono- to few- layer $Bi_2Se_3$ grown on top using chemical vapor deposition.[61–63] However, the same effect has been observed with very thick 10+ quintuple layer $Bi_2Se_3$.[62]

Figure 19a is a schematic showing how patterns can be written in ambient using a laser. Figure 19b shows a change in the color (or optical response) induced by the laser exposure and oxygen intercalation. Figure 19c is a fluorescence image, where only the laser exposed areas illuminate, demonstrating submicron spatial resolution of laser writing, and deterministic tuning of the PL intensity. Figures 19d-e show DFT-relaxed $Bi_2Se_3/MoS_2$ 2D heterostructures in two configurations: (d) pristine, and (e) five oxygen molecules placed in the interlayer region. The calculations predict the 2D heterostructure to be stable, and predict that the intercalated atoms weaken the interlayer coupling. Together, these results demonstrate and variety of promising technologies, including submicron photoluminescent pixels (PLPs), oxygen sensing, ultradense information storage, and ultradense oxygen storage on a chip at ambient.

### 4.4.2. Pressure tuning twist-dependent properties

The hydrostatic pressure surrounding the twisted 2D structure is an experimental knob to tune its phase diagram. More specifically, applying hydrostatic pressure to a 2D structure forces the layers together, thereby decreasing the interlayer separation. This, subsequently, increases the interlayer coupling and facilitates the emergence of new properties, where twist dependent properties associated with the moiré superlattice are likely to get stronger. Applying hydrostatic pressure has been demonstrated to tune a variety of properties, including superconductivity in tBLG,[79] metal-insulator transition in $FePS_3$,[221] Raman modes,[76,222] interlayer excitons,[222] and antiferromagnetic-to-ferromagnetic (AFM-to-FM) transition in $CrI_3$.[75,77]

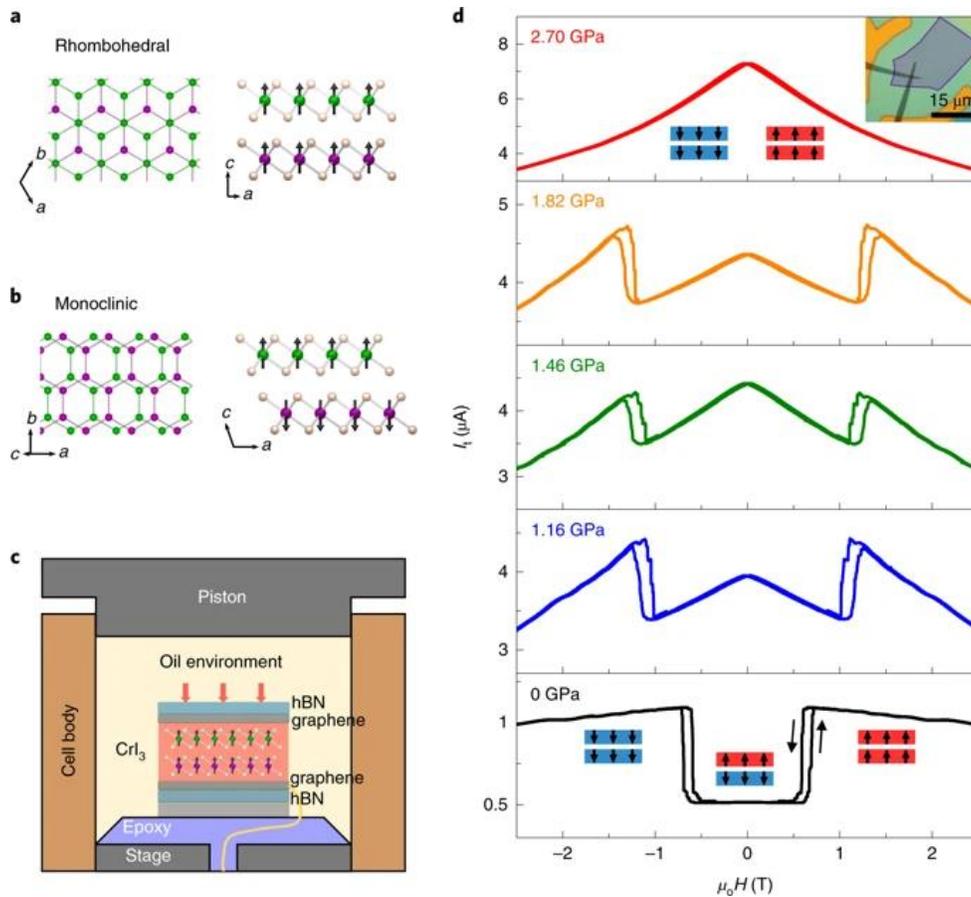

**Figure 20: Pressure tuning the properties of 2D Structures.** (a) Schematic of rhombohedral and (b) monoclinic stacking with the interlayer coupling induced spin directions shown. Rhombohedral contains ferromagnetism, whereas monoclinic contains antiferromagnestism. (c) Schematic of a common high-pressure experimental set-up used in twisted 2D structures. The force applied to the piston exerts a pressure on the device through oil. (d) Tunnelling current versus magnetic field at a series of pressures. Insets: magnetic states (only one of the two time-reversal AFM ground states is shown). Reprinted images and portions of the caption from [75].

## 5. Future Outlook

The high volume of rich science revealed so far makes it adequately evident that moiré superlattice solids can be expected to remain an important source of intellectually-stimulating scientific discoveries for years to come. Interaction within the plane (due to lattice distortions) as well as out of plane (due to charge transfer, orbital hybridization, and distortion of the band structures) are promising ways in which unexpected strong correlations, topological, and symmetry-induced phenomena are going to reveal themselves in what are otherwise well-understood materials. Often, the revealed new phenomena have no trivial relation with the fundamental nature of the parent 2D crystals, and with this point it is impossible to predict how large the volume of unknown new phenomena could be. Given that there are literally hundreds of 2D materials theoretically available to the scientific community, it is reasonable to expect that a significant wealth of quantum phenomena are waiting to be discovered, once high-quality 2D heterostructures stacked with predictable precision become available to a wider range of scientists and engineers.

However, the challenge of investigating all these material-combinations through trial-and-error approaches is also unrealistic, to say the least. To understand this – consider a 2-layered heterostructure, which can be made from 100 different 2D materials with, say, 30 different twist-angles (with 1 degree angular separation). This results in 100x(30*100) = 300,000 possibilities of twisted 2D materials whose electronic, optical, magnetic, or other properties can reveal new science. The actual number could be much larger with larger varieties of 2D materials, possible values of twisting angles, and layer thicknesses. For example, an extension of this analysis to 3 layers results in 100x(30*100)x(30*100) = 9,00,000,000 trilayer possibilities from 100 different 2D materials. While it is very likely that a large number of these possibilities will not reveal transformative new scientific discoveries, even if *one-in-a-million* (or 900) of these combinations lead to non-trivial physical phenomena – this is enough to keep scientists and engineers engaged for the next decade, if not more – if there is a clear pathway for predicting and precisely manufacturing the most likely candidates for these proposed *one-in-a-million* phenomena.

Multi-disciplinary creative and collaborative approaches are needed to overcome this seemingly insurmountable challenge. The first important component is to develop accurate predictive theoretical tools that provide a reasonable starting-point for predicting specific non-trivial phenomena in twisted heterostructures. To this end, a number of groups are exploring the creation of searchable databases for electronic structures of 2D materials. First-principles approaches – such as the density functional theory and its variants are parameter-free and are now being used to systematically build such searchable databases.[223,224] Next, data-driven searches will be needed accelerate the possibility of discovery of target-specific electronic behavior. Such approaches are now becoming more accessible, for example 2D materials for photocatalysis,[225] topologically non-trivial materials,[226] and solar-cell materials.[227] In parallel, new methods need to be developed for precision high-rate synthesis of high-quality 2D materials,[228] and for vertically-stacked 2D materials with high twist-angle precision. Encouraging progress in being made in "search-and-stack" automation already,[194] as such high-volume 2D materials stacking will lead to materials access to a wider range of experimentalists who will be able to expand the scope of discoveries within and beyond the predicted new phenomena. This concerted approach will overcome the current reliance on slow or error-prone human approaches and therefore genuinely open-up the pathway for new levels of intellectual discoveries as well as applications potentials. Further applications of data-mining and pattern-recognition approaches may then fill the gap between the reaches of predictive theory and observed phenomena, thereby enabling more efficient predictions with time. This could genuinely lead to new science, engineering and applications discoveries in atomically-thin materials that are currently beyond the imagination of the community.

Ferromagnet: Fe 3 GeTe 2. *2D Mater.* **2016**, *4* (1), 011005. https://doi.org/10.1088/2053-1583/4/1/011005.

Electronic Grade 2D Materials. *2D Mater.* **2019**, *6* (2), 022001. https://doi.org/10.1088/2053-1583/aaf836.